\documentclass[amsmath, amssymb]{WileyMSP-template}

\usepackage{siunitx}
\usepackage{nicefrac}

\usepackage{setspace}
\DeclareMathAlphabet\mathbfcal{OMS}{cmsy}{b}{n}
\DeclareSIUnit\angstrom{\text {Å}}

\usepackage{gensymb}
\usepackage{xcolor}

\usepackage{amsmath}

\usepackage{ragged2e}

\usepackage{hyperref}
\usepackage{cleveref}
\usepackage[displaymath, mathlines,running]{lineno}
\usepackage{url}

\usepackage{float}

\begin{document}

\pagestyle{fancy}

\title{Probing the Surface Polarization of Ferroelectric Thin Films by X-ray Standing Waves}

\maketitle

\author{Le Phuong Hoang},
\author{Irena Spasojevic},
\author{Tien-Lin Lee},
\author{David Pesquera},
\author{Kai Rossnagel},
\author{Jörg Zegenhagen},
\author{Gustau Catalan},
\author{Ivan A. Vartanyants},
\author{Andreas Scherz, and}
\author{Giuseppe Mercurio*}

\dedication{}

\begin{affiliations}
Le Phuong Hoang\\
European XFEL, 22869 Schenefeld, Germany\\
Max Planck Institute for the Structure and Dynamics of Matter, 22761 Hamburg, Germany\\
Institute of Experimental and Applied Physics, Kiel University, 24098 Kiel, Germany\\
\smallskip
Irena Spasojevic\\
Department de Física, Universitat Autònoma de Barcelona, 08193 Bellaterra, Spain \\
\smallskip
Tien-Lin Lee, Jörg Zegenhagen\\
Diamond Light Source Ltd., Didcot, OX110DE Oxfordshire, United Kingdom\\
\smallskip
David Pesquera\\
Catalan Institute of Nanoscience and Nanotechnology (ICN2), CSIC and BIST, Campus UAB, Bellaterra, 08193 Barcelona, Spain\\
\smallskip
Kai Rossnagel\\
Ruprecht Haensel Laboratory, Deutsches Elektronen-Synchrotron DESY, 22607 Hamburg, Germany\\
Institute of Experimental and Applied Physics, Kiel University, 24098 Kiel, Germany\\
\smallskip
Gustau Catalan\\
Catalan Institute of Nanoscience and Nanotechnology (ICN2), CSIC and BIST, Campus UAB, Bellaterra, 08193 Barcelona, Spain\\
Institucio Catalana de Recerca i Estudis Avançats (ICREA), 08010 Barcelona, Catalonia\\
\smallskip
Ivan A. Vartanyants\\
Deutsches Elektronen-Synchrotron DESY, 22607 Hamburg, Germany\\
\smallskip
Andreas Scherz, Giuseppe Mercurio\\
European XFEL, 22869 Schenefeld, Germany\\
Email: giuseppe.mercurio@xfel.eu\\
\end{affiliations}


\newpage

\begin{justify}

\begin{abstract}
Understanding the mechanisms underlying a stable polarization at the surface of ferroelectric thin films is of particular importance both from a fundamental point of view and to achieve control of the surface polarization itself. In this study, it is demonstrated that the X-ray standing wave technique allows the polarization near the surface of a ferroelectric thin film to be probed directly. The X-ray standing wave technique is employed to determine, with picometer accuracy, Ti and Ba atomic positions near the surface of three differently strained $\mathrm{BaTiO_3}$ thin films grown on scandate substrates, with a $\mathrm{SrRuO_3}$ film as bottom electrode. This technique gives direct access to atomic positions, and thus to the local ferroelectric polarization, within the first 3 unit cells below the surface. By employing X-ray photoelectron spectroscopy, a detailed overview of the oxygen-containing species adsorbed on the surface, upon exposure to ambient conditions, is obtained. The combination of structural and spectroscopic information allows us to conclude on the most plausible mechanisms that stabilize the surface polarization in the three samples under study. The different amplitude and orientation of the local ferroelectric polarizations are associated with surface charges attributed to the type, amount and spatial distribution of the oxygen-containing adsorbates.
\end{abstract}


\section{Introduction}
\label{sec:intro}

Ferroelectric thin films have attracted great scientific interest due to their properties, such as switchable polarization, ferroelasticity, piezoelectricity, and pyroelectricity, which are crucial for technological applications.\textsuperscript{\cite{fernandez_thin-film_2022, khan_future_2020, qi_review_2021, han_ferroelectric_2022}} Displacive ferroelectrics, such as BaTiO$_3$ (BTO), exhibit an intrinsic spontaneous polarization associated with the relative displacement of cations and anions within the unit cell.\textsuperscript{\cite{Franco93}} This polarization can be manipulated by varying the lattice parameters of the ferroelectrics.\textsuperscript{\cite{Choi04, schlom_strain_2007}} To this end, ferroelectric thin films have been grown on substrates with different lattice constants. The lattice mismatch can induce uniform strain or strain gradients in the thin films.\textsuperscript{\cite{Catalan05}} Suitable substrates and bottom electrodes have been employed to tune the ferroelectric polarization, which is typically measured by piezoresponse force microscopy (PFM), a technique that is sensitive to the average polarization of the entire film, for thicknesses of a few tens of nanometer.\textsuperscript{\cite{harnagea_challenges_2004, eng_local_2004}} The distribution of the ferroelectric polarization at the surface, which can differ from that of the bulk, has been investigated theoretically as a function of various parameters, such as surface termination and adsorbates.\textsuperscript{\cite{Gattinoni20, Deleuze20}} However, an experimental method that can simultaneously probe the surface ferroelectric polarization and the chemical composition of the adsorbates is still lacking.

The determination of the surface polarization has a two-fold relevance. First, from a fundamental point of view, uncompensated charges at the surface of a ferroelectric thin film can be screened, among several mechanisms, by external charges provided by adsorbates\textsuperscript{\cite{shin_atomistic_2009, Wang12, lee_imprint_2016, tian_water_2018}} or can lead to a reconstruction of the top unit cells to minimize the surface energy.\textsuperscript{\cite{shin_polar_2008, gao_atomic_2016, domingo_surface_2019}} This, in turn, can influence the polarization of deeper layers and for very thin films affect the polarization of the entire sample.\textsuperscript{\cite{wang_reversible_2009}} Therefore, in order to control the ferroelectric polarization of a thin film, probing it and understanding its stabilization mechanisms near the surface are of particular importance. Second, from the point of view of promising applications, ferrolectrics have been proposed as catalysts with chemical activity that is switchable between reducing and oxidizing surfaces depending on the polarization direction.\textsuperscript{\cite{garrity_chemistry_2010, kakekhani_ferroelectric-based_2015, kakekhani_polarization-driven_2016, vonruti_catalysis_2020, wan_catalysis_2021, lan_enhancing_2021}} In this context, determining the surface polarization is the first step towards the development of efficient ferroelectric catalysts. 

Among non-destructive techniques employed to determine atomic positions at (and near) the surface, and thus the microscopic origin of ferroelectric polarization, crystal truncation rod (CTR) scattering and low-energy electron diffraction (LEED-IV) have been successfully employed to reveal atomic structures with approximately \SI{\pm 10}{\pm} accuracy.\textsuperscript{\cite{shin_atomistic_2009, Wang12, lee_imprint_2016, shin_polar_2008, fong_direct_2005,wang_chemistry_2012}} However, neither of the above methods provides spectroscopic information on atoms in ferroelectric thin films and adsorbate species. This could be achieved by X-ray photoelectron diffraction, however, at the expense of rather complex multiple-scattering simulations.\textsuperscript{\cite{despont_x-ray_2006, despont_direct_2006, pancotti_multiple_2009, pancotti_x-ray_2013, bouwmeester_observing_2022}}

In this work we employ the X-ray standing wave (XSW) technique, a combination of X-ray diffraction (XRD) and X-ray spectroscopy, to determine atomic positions with picometer (pm) accuracy and chemical specificity. The structural accuracy of this technique for determining atomic positions in single crystals and adsorbates on crystal surfaces has been demonstrated extensively.\textsuperscript{\cite{batterman_dynamical_1964, materlik_x-ray_1984, zegenhagen_surface_1993, Woodruff05, Zegenhagen_book_2013}} Furthermore, the XSW technique proved to be successful in determining the polarity of non-centrosymmetric single crystals\textsuperscript{\cite{trucano_use_1976}} and thin films,\textsuperscript{\cite{kazimirov_polarity_1998, kazimirov_high-resolution_2001}} as well as ferroelectric thin films.\textsuperscript{\cite{Bedzyk00, marasco_atomic-scale_2001}} In the latter experiments, the average polarity of the thin films was determined by combining XSW and X-ray fluorescence spectroscopy (XFS). Here, we apply the XSW technique, in combination with  X-ray photoelectron spectroscopy (XPS), a more surface sensitive technique than XFS, to measure the displacement of Ti atoms from the center of the unit cell in differently strained BTO thin films, and thereby deduce the ferroelectric polarization at different depths near the surface. These data are interpreted in the context of the average film polarization measured by PFM, as well as the type, content and spatial distribution of adsorbates on the sample surface. This combination of the structural sensitivity of XSW with the chemical specificity and depth selectivity of XPS, provides a detailed insight into the near-surface polarization distribution and its interplay with adsorbates and the bulk ferroelectric polarization.


\section{X-ray standing waves generated in thin films}
\label{sec:theory}

The XSW technique is particularly useful for determining atomic positions in crystals, surfaces, and their adsorbates.\textsuperscript{\cite{batterman_dynamical_1964, materlik_x-ray_1984, zegenhagen_surface_1993, Woodruff05, Zegenhagen_book_2013}} The interference between incoming and Bragg-diffracted X-ray plane waves in a perfect crystal results in an X-ray standing wave field with the following sinusoidal modulation of the X-ray intensity $I_\mathrm{XSW}$ (\textbf{Figure \ref{fig:setup}a,b}):
\begin{equation}
    I_\mathrm{XSW}\left(E_\nu\right) \propto 1 + \left|\frac{\mathcal{E}_\mathrm{H}}{\mathcal{E}_\mathrm{0}}\right|^2 + 2 \left|\frac{\mathcal{E}_\mathrm{H}}{\mathcal{E}_\mathrm{0}}\right| \cos\left(\alpha(E_\nu) + \boldsymbol{h}\boldsymbol{r} \right),
\label{eq:Ixsw-simple}
\end{equation}
where $\boldsymbol{h} = 2 \pi \boldsymbol{H}$, and $\boldsymbol{H}$ is the reciprocal lattice vector. In Equation \eqref{eq:Ixsw-simple}, the three terms represent the incident, Bragg-diffracted, and interference X-ray wave, respectively. As the incident photon energy $E_\nu$ varies through the (hkl) Bragg reflection, the phase $\alpha(E_\nu)$ between the Bragg-diffracted $\mathcal{E}_\mathrm{H}$ and incident $\mathcal{E}_0$ electric field amplitudes changes by $\pi$. This leads to a shift of the XSW field along $\boldsymbol{H}$ by $d_\mathrm{hkl}/2$, where $d_\mathrm{hkl} = \lvert \boldsymbol{H} \rvert ^{-1}$ is the spacing between two consecutive (hkl) atomic planes (Figure \ref{fig:setup}b). Atoms at different positions in the unit cell experience different X-ray absorption and hence give rise to different photoelectron (PE) yield as a function of the photon energy $E_\nu$. As a result, the atomic positions can be determined with pm spatial resolution by monitoring the corresponding PE yield (Section \ref{sec:results-xsw}).

In a typical ferroelectric thin film grown on a substrate, the lattice mismatch may lead to strain gradients inside the epitaxial layers.\textsuperscript{\cite{Catalan05}} Therefore, thin films are generally characterized by a deformation field $\boldsymbol{u}\left(z\right)$, which defines the actual displacement of atoms from the corresponding position in a perfect crystal, and the static Debye-Waller factor $\mathrm{e}^{-W(z)}$, which accounts for random displacement of atoms from their average position along the $z$ direction (Figure \ref{fig:setup}a). In contrast to the perfect crystal case above, the XSW generated in a thin film is modified by the deformation phase $\varphi\left(z\right) = \boldsymbol{h} \cdot \boldsymbol{u}\left(z\right)$ due to the crystal deformation field. Based on the dynamical theory of diffraction, the Takagi-Taupin equations\textsuperscript{\cite{Takagi62,Takagi69,Taupin64}} describe the propagation of X-rays in a deformed crystal, and thus give the following XSW intensity generated in a typical ferroelectric thin film:
\begin{equation}
I_\mathrm{XSW}\left(E_\nu,z\right) = 1 + R\left(E_\nu,z\right) + 2C \sqrt{R\left(E_\nu,z\right)} \mathrm{e}^{-W\left(z\right)}\cos\left(\alpha(E_\nu,z) + \varphi\left(z\right) + 2\pi z/d_\mathrm{hkl} \right),
 \label{eq:Ixsw-FE}
\end{equation}
where $R_0(E_\nu) = R(E_\nu,0)$ indicates the observable X-ray diffracted intensity at the sample surface ($z=0$), and the parameter $C$ depends on the X-ray polarization\textsuperscript{\cite{Ivan01}} (Section \ref{SM:sec:theory}, Supporting Information). Based on the XSW generated in a ferroelectric thin film, from the PE yield curve
\begin{equation}
\begin{split}
     \kappa^s_\gamma(E_\nu) = I_0^{-1} \int_{0}^{t_{L_0}} dz \rho_\mathrm{yi}\left(E_\nu,z,\gamma\right) |T\left(E_\nu,z\right)|^2 \Big[1 + R\left(E_\nu,z\right) 
     + 2C \sqrt{R\left(E_\nu,z\right)} \mathrm{e}^{-W_0} F^s_{c,\gamma} \\
     \cos\left(\alpha(E_\nu,z) + \varphi_0 + 2 \pi P^s_{c,\gamma} \right) \Big],
\label{eq:XSW-yield}
\end{split}
\end{equation}

we can determine the average position and distribution of atoms $s$, which are defined as the coherent position $P^s_{c,\gamma}$ and coherent fraction $F^s_{c,\gamma}$, respectively. These parameters are equivalent to the phase and amplitude of the structure factor $\boldsymbol{S}^s_{\boldsymbol{h}}  = \sum_{j} \exp(\mathrm{i}\boldsymbol{h} \boldsymbol{r}_j^s ) = \left|\boldsymbol{S}^s_{\boldsymbol{h}}\right| \exp(\mathrm{i} \varphi^s_{\boldsymbol{h},\gamma})$, with atoms at positions $\boldsymbol{r}_j^s$. More specifically, $P^s_{c,\gamma} = \varphi^s_{\boldsymbol{h},\gamma} / (2 \pi)$ and $F^s_{c,\gamma} = |\boldsymbol{S}^s_{\boldsymbol{h}}| \mathrm{e}^{-W_s}$, where $\mathrm{e}^{-W_s}$ is the Debye-Waller factor accounting for thermal and static atomic displacements. Therefore, the absolute average position of atoms $s$ within the unit cell along $\boldsymbol{H}$ is given by $z^s_\gamma = P^s_{c,\gamma} d_\mathrm{hkl}$, and their spatial distribution is characterized by $F^s_{c,\gamma}$, with $0<P^s_{c,\gamma}<1$ and $0<F^s_{c,\gamma}<1$. In particular, $F^s_{c,\gamma} = 1$ refers to all atoms at the same $z$, while $F^s_{c,\gamma} = 0$ corresponds to a uniform distribution of two or more atomic positions across the unit cell.

In Equation \eqref{eq:XSW-yield}, the PE yield function $\kappa^s_\gamma(E_\nu)$ is the sum of yield contributions from atoms in the top layer $L_0$ at positions $0<z<t_{L_0}$, weighted by $\rho_\mathrm{yi}(E_\nu,z,\gamma)$, with XSW transmission $T\left(E_\nu,z\right)$, and normalization factor $I_0 = \int_{0}^{t_{L_0}} dz \rho_\mathrm{yi}(E_\nu,z,\gamma)$. The function $\rho_\mathrm{yi}(E_\nu,z,\gamma) = \exp\left(-z/\lambda_{l,\gamma}\right)$, where $\lambda_{l,\gamma} = \lambda_{l}(E_\nu) \sin \gamma$,  gives the probability of detecting a photoelectron from the atomic core level $l$ at depth $z$, with exit angle $\gamma$ from the sample surface (Figure \ref{fig:setup}c). The parameter $\lambda_{l}\left(E_\nu\right)$ indicates the inelastic mean free path (IMFP), or more correctly the effective attenuation length (EAL) that includes elastic scattering effects.\textsuperscript{\cite{jablonski_electron_2002}} Tuning $\gamma$ allows the surface sensitivity to be varied and provides average atomic positions over different depths from the surface. To determine the average distribution of atoms $s$ at a given exit angle $\gamma$, the measured PE yield curve is fitted with Equation \eqref{eq:XSW-yield} using $P^s_{c,\gamma}$ and $F^s_{c,\gamma}$ as fit parameters. In fact, all other quantities in Equation \eqref{eq:XSW-yield} can be calculated from known sample properties (Section \ref{SM:sec:theory}, Supporting Information), or derived from the fit of Bragg reflectivity data, e.g., the average amorphization $\mathrm{e}^{-W_0}$ of layer $L_0$ and the corresponding deformation phase $\varphi_0$ (Section \ref{sec:results-xrd}).


\section{Results and Discussion}
\label{sec:exp-details}

\subsection{Sample characterization}
\label{sec:samples}

The samples investigated in this work are BTO ferroelectric thin films grown on three different substrates DyScO$_3$ (DSO), GdScO$_3$ (GSO), and SmScO$_3$ (SSO) with a SrRuO$_3$ (SRO) thin film in between serving as the bottom electrode. Sample growth was performed by means of pulsed laser deposition (Section \ref{SM:sec:LPD}, Supporting Information) on substrates with (001) orientation according to the pseudocubic notation.\textsuperscript{\cite{Vailionis11}} The thickness of the BTO and SRO thin films was determined by grazing X-ray reflectivity (Section \ref{SM:sec:XRR}, Supporting Information). The SRO layers have thicknesses in the range of $20-$\SI{26}{\nm}, while the BTO layers are \SI{20}{\nm}, \SI{37}{\nm} and \SI{35.5}{\nm} thick on DSO, GSO and SSO, respectively (Table \ref{tab:thickness-strain}). To determine the average in-plane strain of the thin films, X-ray reciprocal space maps (RSM) around the (-103) substrate Bragg peak were measured (Section \ref{SM:sec:RSM}, Supporting Information). Results reported in Figure \ref{SM:fig:RSM} (Supporting Information) show that all BTO and SRO thin films are coherently strained to their underlying substrates, without any relaxation of the in-plane lattice parameter. The in-plane strain applied by a substrate to the BTO thin film is calculated as $\epsilon^{a}_\mathrm{BTO} = \left(a_\mathrm{BTO} - a_\mathrm{b,BTO} \right) / a_\mathrm{b,BTO}$, by comparing the measured in-plane lattice parameter  of the thin film $a_\mathrm{BTO}$ (Section \ref{SM:sec:RSM}, Supporting Information) with the respective bulk value $a_{b,\mathrm{BTO}} =$ \SI{3.992}{\angstrom}.\textsuperscript{\cite{Franco93}} As a result, the in-plane compressive strain in the BTO thin film is smallest on the SSO substrate ($-0.38\%$, $a_{b,\mathrm{SSO}} =$ \SI{3.977}{\angstrom}\textsuperscript{\cite{setter_ferroelectric_2006}}), it increases on GSO ($-0.63\%$, $a_{b,\mathrm{GSO}} =$ \SI{3.967}{\angstrom}\textsuperscript{\cite{setter_ferroelectric_2006}}), and is largest on DSO ($-1.23\%$, $a_{b,\mathrm{DSO}} =$ \SI{3.943}{\angstrom}\textsuperscript{\cite{setter_ferroelectric_2006}}) (Table \ref{tab:thickness-strain}). The in-plane compressive strain in BTO thin films induces an out-of-plane spontaneous ferroelectric polarization.\textsuperscript{\cite{Choi04}} To determine the average polarization orientation of the as-grown BTO thin films, and to verify that all samples can be electrically switched, piezoresponse force microscopy (PFM) was employed. PFM data (Section \ref{SM:sec:pfm}, Supporting Information) show that the average BTO polarization of BTO/SRO/DSO is down ($\mathrm{P}^\downarrow$), while in the other two samples it is up ($\mathrm{P}^\uparrow$), i.e., with the Ti atom below ($\mathrm{P}^\downarrow$) or above ($\mathrm{P}^\uparrow$) the center of the oxygen octahedra (Figure \ref{fig:setup}b). The different average polarization direction of the three BTO samples may be related to the chemistry and electrostatic potential at the interface between SRO and BTO thin films.\textsuperscript{\cite{Gattinoni20, yu_interface_2012, de_luca_nanoscale_2017}} Finally, both the absence of side peaks in RSM data and the presence of a homogeneous as-grown phase measured by PFM support the presence of a single domain in our samples. This ensures that atomic positions of the same structural phase are measured by XSW measurements.

\begin{table*}[h]
\caption{BTO and SRO layer thicknesses, $t_\mathrm{BTO}$ and $t_\mathrm{SRO}$, resulting from grazing X-ray reflectivity (Section \ref{SM:sec:XRR}, Supporting Information). All substrates are \SI{0.5}{\mm} thick. Compressive in-plane strain in BTO ($\epsilon^{a}_\mathrm{BTO}$) thin films, measured by X-ray reciprocal space maps (Section \ref{SM:sec:RSM}, Supporting Information). Orientation of the average BTO polarization $\mathrm{P}$ measured by PFM (Section \ref{SM:sec:pfm}, Supporting Information).}
\label{tab:thickness-strain}
\begin{tabular}{cccccc}
\hline
  Acronym&Sample&$t_\mathrm{BTO}$ (nm)&$t_\mathrm{SRO}$ (nm)&$\epsilon^{a}_\mathrm{BTO}$ (\%)& $\mathrm{P}$\\ 
 \hline
 BTO/SRO/DSO &  BaTiO$_3$/SrRuO$_3$/DyScO$_3$& 20 & 26 &-1.23& $\downarrow$\\
 BTO/SRO/GSO &  BaTiO$_3$/SrRuO$_3$/GdScO$_3$& 37 & 24 &-0.63& $\uparrow$\\
 BTO/SRO/SSO &  BaTiO$_3$/SrRuO$_3$/SmScO$_3$& 35.5 & 20 &-0.38& $\uparrow$\\
 \hline
\end{tabular}
\end{table*}


\subsection{X-ray standing wave setup}
\label{sec:setup}

XSW experiments were performed at the I09 beamline of Diamond Light Source.\textsuperscript{\cite{Lee18}} The soft X-ray branch of I09, equipped with a plane grating monochromator, delivered an X-ray beam of approximately \SI{300}{\um} $\times$ \SI{200}{\um} full width at half maximum (FWHM) at the sample. Each XSW measurement consists of recording Bragg reflections and simultaneously photoelectron spectra of the sample. The Bragg reflections were measured by scanning the photon energy $E_\nu$ with the incoming X-ray beam impinging on the sample at a fixed angle of incidence $\mathrm{\theta} = 87 \degree$ (Figure \ref{fig:setup}c). The intensity of the diffracted X-ray beam was measured by a Si photodiode with a central through hole for the incident beam to pass (Figure \ref{fig:setup}c). With this experimental geometry, the (001) Bragg reflections of the BTO and SRO films and the substrates were recorded within the range of photon energy $E_\nu$ from \SI{1400}{\eV} to \SI{1700}{\eV}. Simultaneously, XPS spectra were measured by a Scienta EW4000 electron analyzer, with the detection system consisting of a microchannel plate (MCP) followed by a charge-coupled device (CCD). The wide acceptance angle of the electron analyzer enabled parallel measurements of spectra over three different exit angle ranges: $\gamma_1$ ($7.8\degree \pm 5.4\degree$), $\gamma_2$ ($18.5\degree \pm 5.4\degree$) and $\gamma_3$ ($27.4\degree \pm 3.6\degree$). This provides the chemical and structural information of the BTO films with increasing depth sensitivity from $\gamma_1$ to $\gamma_3$. The overall spectral energy resolution, limited by the X-ray bandwidth, was approximately \SI{400}{\meV}. The (001) Bragg reflections, XPS and XSW experimental results are reported in the following Sections \ref{sec:results-xrd}, \ref{sec:results-xps}, and \ref{sec:results-xsw}, respectively.


\subsection{(001) Bragg reflections}
\label{sec:results-xrd}

To determine the atomic positions at the BTO surface along the out-of-plane polarization direction, the (001) Bragg reflection was chosen for the XSW experiments. \textbf{Figure \ref{fig:xrd-c}a} shows the reflectivity curves around the (001) reflections of the BTO, SRO, and substrates. The reflections of the substrate are more than two orders of magnitude stronger and much narrower compared to those of the thin layers. The low intensity and the broadening of the thin film Bragg peaks results from two factors: the finite film thickness and the inhomogeneous strain. More specifically, the reflectivity of BTO [SRO] films lies in the range of $0.03\% - 0.05\%$ [$0.01\% - 0.02\%$], whereas the substrate reflectivity is of the order of $10\%$ for all samples. Going from SSO, to GSO, and then to DSO, the substrate Bragg peak gradually shifts to higher photon energies, as expected from the decreasing trend of the bulk out-of-plane lattice parameters of the substrates (Figure \ref{SM:fig:RSM}, Supporting Information). This trend suggests that the in-plane strain in the films becomes more compressive in BTO and less tensile in SRO (Section \ref{SM:sec:SRO_structure}, Supporting Information) in the same substrate order. As a result, the average $c$ parameters of BTO (Section \ref{SM:sec:c_average}, Supporting Information) increase from the SSO to the DSO sample: $\overline{c}_\mathrm{BTO} = $ \SI{4.055 \pm 0.022}{\angstrom}, \SI{4.063 \pm 0.023}{\angstrom}, \SI{4.070 \pm 0.045}{\angstrom}, respectively. For the BTO thin films, the measured $\overline{c}_\mathrm{BTO}$ parameters correspond to an average out-of-plane strain $\overline{\epsilon}^c_\mathrm{BTO} = \left( \overline{c}_\mathrm{BTO} - c_{b,\mathrm{BTO}} \right) / c_{b,\mathrm{BTO}}$ of $0.48\%$, $0.68\%$ and $0.84\%$ for SSO, GSO and DSO, respectively, with the bulk out-of-plane lattice parameter $c_{b,\mathrm{BTO}} =$ \SI{4.036}{\angstrom}.\textsuperscript{\cite{Franco93}}

As anticipated in Section \ref{sec:theory}, an epitaxial thin film may be characterized by an inhomogeneous out-of-plane strain. According to the general strain profile model discussed in Refs.\textsuperscript{\cite{Catalan05,kim_thermodynamic_1999}}, the strain gradient is expected to be proportional to the strain, ${\partial \epsilon}^c/{\partial z} \propto \epsilon^c$, independently of the actual relaxation mechanism. As a result, the out-of-plane parameter $c$ follows an exponential dependence on $z$:
\begin{equation}
    c(z) = c_b \left(1 + \epsilon_\mathrm{int}^c \mathrm{e}^{-(t-z)/\delta} \right),
\label{eq:c-gradient}
\end{equation}
where $\epsilon_\mathrm{int}^c = (c_\mathrm{int} -c_b)/c_b$ is the strain at the interface with the underlying layer (or substrate), and $\delta$ is the penetration depth of strain that is inversely proportional to the strain gradient. This model has been successfully applied to ferroelectric thin films of several 100s \SI{}{\nm}.\textsuperscript{\cite{Catalan05,kim_thermodynamic_1999}} In a recent work,\textsuperscript{\cite{lichtensteiger_interactivexrdfit_2018, weymann_full_2020}} it was also found that \SI{50}{\nm} thick $\mathrm{PbTiO_3}$ films, displaying high crystalline quality and no indication of in-plane relaxation, could be well described by an exponential profile of $c$ parameters, as further confirmed by TEM images. In this case,\textsuperscript{\cite{weymann_full_2020}} the strain gradient was assigned to a compositional gradient of lead oxide dipolar vacancies. A similar distribution of vacancies or defects could be present also in our samples and may underlay the presence of strain gradients. In fact, in our work modeling BTO and SRO thin films with a constant $\overline{c}$ leads to unsatisfactory fit results. To fit the experimental reflectivity curves in Figure \ref{fig:xrd-c}, BTO and SRO layers are divided into $n$ sublayers $L_i$ (with $i = 0, ..., n-1$) of equal thickness $t_i$ with an out-of-plane lattice parameter $c_i$ varying exponentially with $i$ as described by Equation \eqref{eq:c-gradient}, Debye-Waller factor $\mathrm{e}^{-W_i}$, and deformation phase $\varphi_i = 2 \pi ( c_i - \overline{c})t_i/\overline{c}^2$ (Section \ref{SM:sec:deformation_phase}, Supporting Information). In our samples, the minimum common number of sublayers necessary to accurately describe them is $n=5$. Increasing the number of sublayers $n$ does not improve the fit. Experimental data in Figure \ref{fig:xrd-c}a are fitted with the reflectivity $R_0(E_\nu)$ using the fitting parameters $\mathrm{e}^{-W_i}$ ($i=0, ...,4$), $\epsilon_\mathrm{int}^c$ and $\delta$ for the BTO and SRO layer (Section \ref{sec:theory} and Section \ref{SM:sec:theory}, Supporting Information). 

The best fits to the reflectivity curves shown in Figure \ref{fig:xrd-c}a reproduce reasonably well the main features of the experimental data. This validates the strain gradient theory used to model these samples. The resulting out-of-plane parameters $c_i$ are shown in Figure \ref{fig:xrd-c}b and in Figure \ref{SM:fig:c_SRO} (Supporting Information). The Debye-Waller factors $\mathrm{e}^{-W_i}$ in BTO sublayers are mostly $\geq 0.9$ with lower values at the interface to SRO (Table \ref{SM:tab:reflfit_W}, Supporting Information). A similar observation of larger structural disorder at the interface to the layer below has been revealed by other transmission electron microscopy (TEM) studies.\textsuperscript{\cite{borisevich_suppression_2010, kim_direct_2014, wu_direct_2020}} The fit parameters $\epsilon_\mathrm{int}^c$ and $\delta$ of BTO show a direct and inverse proportionality to the in-plane compressive strain $\epsilon^{a}_\mathrm{BTO}$, respectively (Table \ref{SM:tab:reflfit_eps_delta}, Supporting Information). In fact, a larger in-plane compressive strain $\epsilon^a_\mathrm{BTO}$ leads to larger average out-of-plane strain $\overline{\epsilon}^c_\mathrm{BTO}$ and therefore, a larger strain gradient up to ${\partial \epsilon^c}/{\partial z} =$ \SI{1.7e-3}{\per\nm} in BTO/SRO/DSO. Similar and even larger strain gradients have been previously observed by TEM and CTR scattering measurements on BTO thin films\textsuperscript{\cite{lee_imprint_2016, wu_direct_2020}} and other ferroelectrics.\textsuperscript{\cite{weymann_full_2020,borisevich_suppression_2010, kim_direct_2014}} Interestingly, despite the different in-plane compressive strains in the three samples (Table \ref{tab:thickness-strain}), the strain gradients lead to similar out-of-plane lattice parameters $c_0$ at the top sublayer $L_0$: \SI{4.038}{\angstrom}, \SI{4.046}{\angstrom}, \SI{4.045}{\angstrom}. Subsequently, the fitting of the experimental reflectivity curves in Figure \ref{fig:xrd-c}a provides the necessary structural data to calculate the PE yield fit function $\kappa^s_\gamma(E_\nu)$ in Equation \eqref{eq:XSW-yield}. We turn now to the determination of the experimental PE yield from XPS spectra.


\subsection{Ba and Ti XPS}
\label{sec:results-xps}

The photoelectron yield $\kappa^s_\gamma(E_\nu)$ of atomic species $s$, measured at exit angle range $\gamma$ and photon energy $E_\nu$, is defined as the corresponding PE peak integral after background subtraction. To determine the PE yield of Ba and Ti atoms in the BTO thin films, Ba 4d and Ti 2p PE spectra were measured over the three exit angle ranges ($\gamma_1$, $\gamma_2$, $\gamma_3$), and the results are reported in \textbf{Figure \ref{fig:XPS-Ba-Ti}}. The Ti spectra in Figure \ref{fig:XPS-Ba-Ti}a show the 2p doublet at $458.8$ and \SI{464.5}{\eV}. Spectra measured at different exit angles show the same spectral shape. This indicates that Ti atoms at different depths from the BTO surface experience the same chemical environment for the formation of the nominal Ti$^{4+}$ state. In particular, the absence of a peak at \SI{1.7}{\eV} below the main Ti 2p$_{3/2}$ peak provides evidence for the absence of oxygen vacancies leading to Ti$^{3+}$ near the BTO surface.\textsuperscript{\cite{Wang12}} In contrast, Ba 4d spectra, displayed in Figure \ref{fig:XPS-Ba-Ti}b, show at least two kind of atomic species. The spin-orbit split levels Ba 4d$_{5/2}$ and Ba 4d$_{3/2}$ at \SI{88.8}{\eV} and at \SI{91.4}{\eV} originate from the Ba atoms below the surface and are hence referred to as the bulk component ($\mathrm{Ba_{bulk}}$). On the other hand, the $\mathrm{Ba_{surf}}$ peaks exhibit a binding energy shift $\Delta \mathrm{BE}=+$\SI{1.2}{\eV} and twice the FWHM compared to the bulk component. The enhancement of $\mathrm{Ba_{surf}}$ in the most surface sensitive spectrum (Ba($\gamma_1$)) clearly indicates its correspondence to Ba atoms at the BTO surface. The assignment of the BE-shifted Ba components is still controversial, with some studies attributing them to BaCO$_3$ or $\mathrm{Ba(OH)_2}$,\textsuperscript{\cite{wang_chemistry_2012, kumar_investigations_2003, wang_chemistry_2014}} and others to BaO$_2$ species.\textsuperscript{\cite{droubay_work_2015, Irena21}} Regardless of the chemical binding of Ba atoms, a recent X-ray photoelectron diffraction study revealed that these Ba atoms are only located at the topmost BaO plane.\textsuperscript{\cite{deleuze_nature_2022}} Our data confirm the surface origin of the BE-shifted Ba 4d component, and the larger FWHM is consistent with multiple chemical environments surrounding the surface Ba atoms observed in previous work. As a result, the PE yield of the surface unit cell ($\gamma_1$) is given by the sum of the two components ($\mathrm{Ba_{surf}} + \mathrm{Ba_{bulk}}$) in order to include both Ba atoms at the surface and just below it, while the Ba PE yield of the deeper unit cells (at $\gamma_2$ and $\gamma_3$) is given only by the $\mathrm{Ba_{bulk}}$ component. On the other hand, the Ti PE yield is given by the total area of the Ti doublet for all exit angle ranges. As shown above, the possibility provided by XPS to distinguish between atoms in different chemical environments enables the XSW technique to selectively determine the positions of different chemical species of the same element.


\subsection{Ba and Ti XSW}
\label{sec:results-xsw}

\textbf{Figure \ref{fig:XSW-data}} shows the normalized PE yields of Ti and Ba, $\kappa^{\mathrm{Ti}}_\gamma(E_\nu)$ and $\kappa^{\mathrm{Ba}}_\gamma(E_\nu)$, measured over the exit angle ranges $\gamma_1$, $\gamma_2$, and $\gamma_3$ around the BTO (001) Bragg peak of the three samples under study. Each $\kappa^s_\gamma(E_\nu)$ and corresponding error bar $\sigma_\kappa$ result from the average and standard deviation of $N$ photoelectron yield profiles ($5<N<10$) measured under the same conditions. Each PE yield curve shown in Figure \ref{fig:XSW-data} is normalized by the intensity of the incoming X-ray beam and by the respective photoionization cross-section over the measured photon energy range (Section \ref{SM:sec:yield-normalization}, Supporting Information). In correspondence to the BTO reflectivity maxima $\kappa^{\mathrm{Ti}}_\gamma(E_\nu)$ curves show a peak-like shape, while $\kappa^{\mathrm{Ba}}_\gamma(E_\nu)$ profiles display a dip. This can be explained as follows. For the BTO (001) reflection, the Bragg diffraction planes are near the Ba atomic planes\textsuperscript{\cite{bedzyk_probing_2000, bedzyk_two-beam_1985}} (dashed lines in Figure \ref{fig:setup}b). When the incoming X-ray photon energy reaches the Bragg condition ($E_\nu \approx E_B$) from the low-energy side, the XSW forms with a sinusoidal modulation of the X-ray intensity $I_{\mathrm{XSW}}$ and period $d_{001}$ along $\boldsymbol{H}$. At this point, the standing wave antinodes and nodes are between and at the diffraction planes, respectively (Figure \ref{fig:setup}a,b). Therefore, Ti atoms, which are nearly half way between two adjacent diffraction planes and hence more aligned with the antinodes, show an increase in the PE yield, while Ba atoms (near the diffraction planes and aligned with the nodes) experience a decrease in $I_\mathrm{XSW}$ and consequently have smaller $\kappa^{\mathrm{Ba}}_\gamma(E_\nu)$. As the photon energy is varied through the Bragg condition ($E_\nu>E_B$), the nodes and antinodes move by $d_{001}/2$ along $\boldsymbol{H}$ and the XSW intensity modulation fades away. Because of the weak diffraction of the incoming X-ray wave from the thin film, the reflectivity maxima of our samples range from $0.02\%$ to $0.05\%$. Therefore, the interference between the incoming and Bragg-diffracted X-ray waves results in a weak XSW intensity modulation with an amplitude, which is proportional to $2\sqrt{R_0(E_\nu)}$, of less than $4\%$. Nevertheless, as shown below, this is sufficient to determine, from the information encoded in the PE yield profiles, the average atomic distribution within the unit cell with pm spatial accuracy.

\begin{table*}[b]
\caption{Coherent position $P_{c, \gamma}^s$ and coherent fraction $F_{c, \gamma}^s$ of Ba and Ti PE yield fits (Figure \ref{fig:XSW-data}) at the exit angle ranges $\gamma_1$, $\gamma_2$, and $\gamma_3$, in the three samples under study. Coherent position offset of Ti atoms from the center of the unit cell defined by Ba atoms, $\Delta P^{\mathrm{Ti}}_{c,\gamma} = P^{\mathrm{Ti}}_{c,\gamma} - \left(P^{\mathrm{Ba}}_{c,\gamma}-0.5\right)$, and absolute off-center displacement of Ti atoms, $\Delta z^{\mathrm{Ti}}_\gamma = c_0 \Delta P^{\mathrm{Ti}}_{c,\gamma}$, expressed in pm.}
\begin{tabular}{cc|cc|cc|cc}
\hline
 Sample & Angle range & $P_{c, \gamma}^{\mathrm{Ba}}$ &$F_{c, \gamma}^{\mathrm{Ba}}$ & $P_{c, \gamma}^{\mathrm{Ti}}$ & $F_{c, \gamma}^{\mathrm{Ti}}$ & $\Delta P^{\mathrm{Ti}}_{c,\gamma}$ & $\Delta z^{\mathrm{Ti}}_\gamma$ (pm)\\ 
 \hline
 & $\gamma_1$ & 0.99(1) & 0.47(4) & 0.54(3) & 0.34(7) & 0.05(5)&20\\
 BTO/SRO/DSO& $\gamma_2$ & 1.00(1) & 0.81(5) & 0.54(1) & 0.55(6)& 0.04(2)&16\\
 & $\gamma_3$ & 1.00(1) & 0.82(5) & 0.52(3) & 0.46(8)& 0.02(4)&8\\
 \hline 
 & $\gamma_1$& 1.05(1) & 0.46(4) & 0.54(3) & 0.29(6) & -0.01(4)&-4 \\
 BTO/SRO/GSO& $\gamma_2$& 1.06(1) & 0.88(3) & 0.55(2) & 0.32(5) & -0.01(3)&-4\\
 & $\gamma_3$& 1.04(1) & 0.78(3) & 0.56(1) & 0.54(5) & 0.02(2)&8\\
 \hline 
 & $\gamma_1$& 1.08(2) & 0.42(7) & 0.63(6) & 0.44(19) & 0.05(8)&20\\
BTO/SRO/SSO& $\gamma_2$& 1.02(1) & 1.00(3) & 0.55(4) & 0.41(12) & 0.03(5)&12\\
& $\gamma_3$& 1.02(1) & 1.00(7) & 0.56(2) & 0.59(9) & 0.04(3)&16\\
\hline
\end{tabular}
\label{tab:Pc-Fc}
\end{table*}

The XSW analysis presented here is based on the calculation of the reflectivity $R_0(E_\nu)$ which assumes either upward or downward average polarization of the BTO film, as it results from PFM data (Table \ref{tab:thickness-strain}). The respective positions of Ba, Ti and O atoms in the unit cell for the reflectivity calculations come from known BTO bulk values.\textsuperscript{\cite{Franco93}} The model employed is validated by the reasonably good fit of both reflectivity and yield data. In fact, the experimental PE yields shown in Figure \ref{fig:XSW-data} are well fitted by Equation \eqref{eq:XSW-yield} with the fitting results, $P^s_{c,\gamma}$ and $F^s_{c,\gamma}$, summarized in Table \ref{tab:Pc-Fc}. As expected from the atomic coordinates used to construct the structural model in the XSW analysis, Ba atoms have $P^{\mathrm{Ba}}_{c,\gamma}\approx1$, while Ti atoms have $P^{\mathrm{Ti}}_{c,\gamma}\approx0.5$. Their exact atomic positions vary with sample and depth by up to few tens of pm with an error bar (averaged over $\gamma$) of \SI{4}{\pm} for Ba and \SI{12}{\pm} for Ti (Section \ref{SM:sec:uncertainty}, Supporting Information). In the context of a displacive ferroelectric like BTO,\textsuperscript{\cite{Zhang22}} the relevant physical quantity is the displacement of Ti atoms from the center of the unit cell (defined by $P^{\mathrm{Ba}}_{c,\gamma}$), which directly relates to the ferroelectric polarization.\textsuperscript{\cite{abrahams_atomic_1968}} Therefore, we calculate $\Delta P^{\mathrm{Ti}}_{c,\gamma} = P^{\mathrm{Ti}}_{c,\gamma} - \left(P^{\mathrm{Ba}}_{c,\gamma}-0.5\right)$ and the corresponding absolute off-center displacement (in pm) of Ti atoms $\Delta z^{\mathrm{Ti}}_\gamma = c_0 \Delta P^{\mathrm{Ti}}_{c,\gamma}$. XSW data at different exit angle ranges $\gamma$ provide Ti atomic displacements at different depths $z$ from the BTO surface. For Ba 4d [Ti 2p] photoelectrons at the (001) BTO Bragg energy $E_B$, the IMFPs are $\lambda_{l,\gamma} =$ \SI{3.4}{\angstrom}, \SI{8.0}{\angstrom}, \SI{11.6}{\angstrom} [\SI{2.7}{\angstrom}, \SI{6.4}{\angstrom}, \SI{9.2}{\angstrom}] for $\gamma_1$, $\gamma_2$ and $\gamma_3$, respectively. The corresponding probability yield functions $\rho_\mathrm{yi}(z)$, reported in \textbf{Figure \ref{fig:sketch}a}, indicate that $\Delta z^{\mathrm{Ti}}_{\gamma_1}$ relates mostly ($\approx 70\%$) to atoms within the first unit cell, while $\Delta z^{\mathrm{Ti}}_{\gamma_2}$ [$\Delta z^{\mathrm{Ti}}_{\gamma_3}$] results primarily from a $\rho_\mathrm{yi}(z)$-weighted average of atomic positions within the top 2 [3] unit cells. Ti atomic displacements in the top three BTO unit cells resulting from the respective $\Delta z^{\mathrm{Ti}}_{\gamma}$ are displayed in Figure \ref{fig:sketch}b-d (Table \ref{tab:Pc-Fc}). In particular, the XSW fit results of the BTO/SRO/DSO sample reveal positive $\Delta z^{\mathrm{Ti}}_\gamma$ values which decrease as $\gamma$ increases. This corresponds to an upward ferroelectric polarization $\mathrm{P}^\uparrow$ with decreasing amplitude from the surface to the bulk. In contrast, the BTO/SRO/GSO sample shows a minor offset $\Delta z^{\mathrm{Ti}}_\gamma < 0$ for $\gamma_1$ and $\gamma_2$, while $\Delta z^{\mathrm{Ti}}_{\gamma_3} > 0$. This indicates an upward polarization $\mathrm{P}^\uparrow$ at larger depths that nearly vanishes with a minor reversal just below the surface. Finally, in the BTO/SRO/SSO sample, $\Delta z^{\mathrm{Ti}}_{\gamma} > 0$, i.e., an upward polarization ($\mathrm{P}^\uparrow$), is observed for all $\gamma$. 

We now move to discussing the atomic coherent fractions. For Ba at the exit angle ranges $\gamma_2$ and $\gamma_3$ the coherent fraction is relatively high in all samples ($F^{\mathrm{Ba}}_{c,\gamma} > 0.8$), indicating high structural order. In particular, in BTO/SRO/SSO, where no reversal of ferroelectric polarization with depth is observed, $F^{\mathrm{Ba}}_{c,\gamma_2}$ and $F^{\mathrm{Ba}}_{c,\gamma_3}$ are equal to $1$. The latter values are an overestimation because at room temperature atomic vibrations lead to $F_{c} < 1$, even in a perfectly ordered atomic layer. The overestimation of $F_{c}$ is due to two possible reasons. First, our XSW analysis does not include non-dipolar parameters, which are currently not available for p, d, and f initial states,\textsuperscript{\cite{vartanyants_non-dipole_2005}} and thus, higher $F_{c}$ are expected without correcting for non-dipole effects. Second, the nonlinear behaviour of the MCP may lead to an overestimation of the count rate, consequently of the XSW modulation amplitude, and thus of $F_{c}$. Conversely, in BTO/SRO/DSO and BTO/SRO/GSO, $F^{\mathrm{Ba}}_{c,\gamma_2}$ and $F^{\mathrm{Ba}}_{c,\gamma_3}$ values are $12\%$ to $22\%$ lower, depending on sample and $\gamma$. This is attributed to the averaging over atoms in unit cells with different polarizations, which also contribute to the generally lower coherent fraction of Ti atoms in the range of $0.3 - 0.6$. On the other hand, for the most surface sensitive measurements at $\gamma_1$, both Ba and Ti atoms have a lower coherent fraction of $0.3$ to $0.5$. The generally lower $F^{s}_{c,\gamma}$ at the surface can be attributed to the larger structural disorder induced by the interactions of the atoms at the topmost oxide plane with adsorbates.


\subsection{Discussion}
\label{sec:discussion}

XSW data reveal that the absolute displacements of Ti atoms from the center of the unit cell ($\Delta z^{\mathrm{Ti}}_{\gamma_i}$) decrease from BTO/SRO/SSO, through BTO/SRO/DSO, to BTO/SRO/GSO. This trend is not correlated with the in-plane compressive strain $\epsilon^{a}_\mathrm{BTO}$, as it could have been expected. Instead, we explain the measured off-center displacements of Ti atoms in light of oxygen-containing adsorbates at the surface, as detailed below. Upon exposure to ambient conditions water adsorbs on the BTO surface and dissociates into OH$^-$ and H$^+$. While OH$^-$ chemisorbs on top of cations (Ba or Ti) or at O vacancies, H$^+$ binds to a lattice oxygen atom ($\mathrm{O_L}$) at or below the surface to form $\mathrm{O_LH^-}$.\textsuperscript{\cite{Wang12, lee_imprint_2016, park_effect_2000}} Our depth-dependent O1s XPS spectra (Section \ref{SM:sec:results-O1s-xps}, Supporting Information) reveal the presence of: (i) negatively charged chemisorbed O species, i.e., $\mathrm{OH}^-$ (hydroxyl groups) or $\mathrm{O_2}^-$ (peroxo groups), modeled by the component $\mathrm{O(4)}$, and (ii) $\mathrm{O_LH^-}$ species, resulting from a $\mathrm{H^+}$ ion bound to an $\mathrm{O_L}$, or a hydroxyl group adsorbed at a oxygen-vacancy site, modeled by the component $\mathrm{O(2)}$. Negatively charged $\mathrm{OH}^-$ or $\mathrm{O_2}^-$ molecules favor the upward polarization $\mathrm{P}^\uparrow$,\textsuperscript{\cite{Gattinoni20}} while positively charged $\mathrm{H^+}$ atoms of $\mathrm{O_LH^-}$ species favor the downward polarization $\mathrm{P}^\downarrow$.\textsuperscript{\cite{shin_atomistic_2009}}

In the BTO/SRO/DSO sample, the BTO film has downward average polarization $\mathrm{P}^\downarrow$ (Table \ref{tab:thickness-strain}) with upward polarization $\mathrm{P}^\uparrow$ in the topmost unit cells (Figure \ref{fig:sketch}b). The latter is favored by negatively charged $\mathrm{OH}^-$ or $\mathrm{O_2}^-$ molecules adsorbed on the surface, represented by component $\mathrm{O(4)}$ in Figure \ref{SM:fig:xps-O1s}a (Supporting Information). In particular, the off-center displacement of Ti atoms decreases from the surface towards the bulk. This trend is consistent with a reversal of the ferroelectric polarization below the top three unit cells, which however is beyond our XSW depth sensitivity. For this configuration to be stable, a concentration of positive charges at the polarization flip interface is required. Importantly, the depth dependence of component $\mathrm{O(2)}$ in this sample is consistent with a distribution of $\mathrm{O_LH^-}$ species over about $4$ unit cells ($\approx$ \SI{15}{\angstrom}) below the surface (Figure \ref{SM:fig:xps-O1s}d, Supporting Information), and thus indicates the accumulation of $\mathrm{H}^+$ atoms as a possible charge compensation mechanism for the polarization reversal below the surface (Figure \ref{fig:sketch}b). A similar scenario has been suggested by Lee and coworkers.\textsuperscript{\cite{lee_imprint_2016}} In their study, atomic positions across the BTO thin film were derived from CTR scattering experiments, while the increase in component $\mathrm{O(2)}$ upon water adsorption was assigned to the presence of $\mathrm{H}^+$ or defects. In our study, the combination of depth-dependent XSW and XPS, and PFM provides further experimental evidence that suggests the presence of a $\mathrm{H}^+$-mediated polarization reversal below the BTO surface.

In the BTO/SRO/GSO sample, the BTO film has an upward average polarization $\mathrm{P}^\uparrow$ (Table \ref{tab:thickness-strain}) in agreement with a positive displacement of Ti atoms in the most bulk-sensitive data ($\Delta z^{\mathrm{Ti}}_{\gamma_3} > 0$). In contrast, the more surface-sensitive data show a small negative displacement of Ti atomic positions (Figure \ref{fig:sketch}c), which indicates a minor polarization reversal at the top unit cells. The polarization reversal at the surface of a $\mathrm{P}^\uparrow$-polarized BTO thin film, upon hydroxylation of a BaO-terminated surface with $\mathrm{O_LH^-}$, has been predicted by DFT calculations and supported by LEED-IV data.\textsuperscript{\cite{shin_atomistic_2009}} In the present work, we go one step further and simultaneously provide structural and the following chemical evidence in support of this predicted behaviour. In fact, in this sample, contrary to the one above, O1s XPS data show that $\mathrm{O_LH^-}$ species are confined at the surface with concentration similar to $\mathrm{OH}^-$ or $\mathrm{O_2}^-$ species (Figure \ref{SM:fig:xps-O1s}b,d, Supporting Information). The minor displacement of Ti atoms from the center of the unit cells near the surface, resulting into a vanishing net polarization, is attributed to the competition between positively charged $\mathrm{H^+}$ atoms of $\mathrm{O_LH^-}$ species that favor downward polarization $\mathrm{P}^\downarrow$ and negatively charged $\mathrm{OH^-}$ or $\mathrm{O_2}^-$ molecules that favor upward polarization $\mathrm{P}^\uparrow$.\textsuperscript{\cite{Gattinoni20}}

In the BTO/SRO/SSO sample, the BTO film has an upward average polarization $\mathrm{P}^\uparrow$ (Table \ref{tab:thickness-strain}), and a positive off-center displacement of Ti atoms throughout the top BTO unit cells is observed (Figure \ref{fig:sketch}d). This scenario implies upward polarization $\mathrm{P}^{\uparrow}$ with accumulation of positive bound charge at the surface. To stabilize this configuration, a compensating negative screening charge at the surface is required. O 1s XPS spectra (Figure \ref{SM:fig:xps-O1s}c, Supporting Information) show a large concentration of negatively charged O species (OH$^-$ and/or $\mathrm{O_2}^-$). In this context, the larger displacement of Ti atoms at the surface (\SI{20}{\pm}), as compared to deeper unit cells, can be understood as a direct influence of dissociated water molecules on BTO surface atoms. A recent DFT study\textsuperscript{\cite{Deleuze20}} predicted that in TiO$_2$-terminated samples OH$^-$ binds to Ti atoms at the surface, thereby inducing a larger offset of Ti atoms in the topmost atomic plane. Our XSW data provide direct experimental evidence for this predicted behavior. Moreover, further oxidation of the hydroxyl group Ti-OH leads to the formation of different peroxide species (e.g., $\mathrm{Ti-O^-}$, $\mathrm{Ti=O_2^-}$)\textsuperscript{\cite{domingo_water_2019}} with negatively charged O atoms that contribute to the component $\mathrm{O(4)}$ and favor $\mathrm{P}^{\uparrow}$ polarization,\textsuperscript{\cite{Gattinoni20,Deleuze20}} as shown by our experimental data.

In single-domain BTO bulk crystals at room temperature with upward $\mathrm{P}^{\uparrow}$ [downward $\mathrm{P}^{\downarrow}$] polarization,
neutron diffraction analysis revealed that the displacement of Ti atoms is \SI{5}{\pm} above [below] the center of the unit cell.\textsuperscript{\cite{Franco93}} In comparison, the most bulk-sensitive XSW data $\Delta z^{\mathrm{Ti}}_{\gamma_3}$ show larger Ti atomic displacements. This can be explained by the residual strain at the top sublayer, leading to out-of-plane lattice parameters $c_0$ larger than the bulk value $c_{b,\mathrm{BTO}}$, and consequently larger atomic displacements. Moreover, as shown above, atomic positions near the surface are influenced by adsorbates, which may lead to smaller (BTO/SRO/GSO) or larger (BTO/SRO/DSO and BTO/SRO/SSO) atomic displacements depending on their type and content.

In summary, the three samples under study have different in-plane compressive strain and thickness (Table \ref{tab:thickness-strain}), however the corresponding strain gradients lead to a similar average out-of-plane lattice parameter $c_0$ in the top sublayer of the BTO/SRO/GSO and BTO/SRO/SSO samples and a slightly smaller one in the BTO/SRO/DSO sample. Upon exposure to ambient conditions each sample displays a different distribution of the ferroelectric polarization near the surface with the common result of screening the bulk polarization and stabilizing the ferroelectric thin film surface. The available data show that there is a correlation between the local polarization at the top unit cells and the type and content of adsorbates on the surface. The interplay of available adsorbates and bulk ferroelectric polarization leads to the resulting distribution of local polarization near the surface. Further studies are required to elucidate to which extent the adsorption of external species influence or are influenced by the polarization below the surface.


\section{Conclusions}

In this work, the XSW technique is successfully employed to BTO thin films to determine the displacement of Ti atoms from the center of the unit cell, defined by Ba atomic positions. In previous studies, the XSW technique in combination with XFS has been employed to determine the polarization orientation of the entire film ($\mathrm{P}^\uparrow$ or $\mathrm{P}^\downarrow$).\textsuperscript{\cite{kazimirov_polarity_1998, kazimirov_high-resolution_2001, marasco_atomic-scale_2001, bedzyk_probing_2000, kazimirov_x-ray_1992, kazimirov_excitation_1997, kazimirov_x-ray_2000, kazimirov_xray_2004}} Here, we have measured the photoelectron yield to determine the near-surface displacement of Ti atoms independently from the polarization in the bulk of the thin film. First, modeling X-ray diffraction data has provided the distribution of out-of-plane lattice parameters resulting from the epitaxial strain in our thin films. Second, the structural sensitivity of the X-ray standing wave combined with the chemical specificity, surface sensitivity and depth selectivity of X-ray photoelectron spectroscopy has provided Ti and Ba atomic positions at different depths with pm spatial resolution. Since the Ba position defines the center of the unit cell, the measure of the Ti position gives direct access to the local ferroelectric polarization at (and near) the surface. The ferroelectric polarization in the top unit cells of the BTO samples under study has been interpreted with the help of depth-dependent O 1s XPS spectra. A detailed analysis of oxygen species adsorbed on the surface has suggested the possible charge compensation mechanisms that are consistent with the distribution of ferroelectric polarizations derived from XSW data. In particular, we have identified three different scenarios: (i) a polarization reversal from downward $\mathrm{P}^\downarrow$ to upward $\mathrm{P}^\uparrow$ polarization leading to a tail-to-tail polarization configuration near the third unit cell that could be stabilized by $\mathrm{H^+}$ atoms diffused below the surface (Figure \ref{fig:sketch}b); (ii) a minor polarization reversal from upward $\mathrm{P}^\uparrow$ to downward $\mathrm{P}^\downarrow$ polarization above the third unit cell leading to a vanishing polarization, as a result of the competing presence of $\mathrm{O_LH^-}$ which favors downward polarization $\mathrm{P}^\downarrow$ and $\mathrm{OH^-}$ or $\mathrm{O^{2-}}$ which favor upward polarization $\mathrm{P}^\uparrow$ (Figure \ref{fig:sketch}c); (iii) a uniform upward polarization $\mathrm{P}^\uparrow$ in the BTO film up to the surface accompanied by a large concentration of compensating negatively charged chemisorbed oxygen species (Figure \ref{fig:sketch}d).

The variety of the observed ferroelectric configurations demonstrates the complexity of these systems and underscores the importance of investigating them from different perspectives to gain a better understanding of the mechanisms that determine the ferroelectric polarization at the surface. The novelty of this work lies in the combination of structural and spectroscopic information, offered by the XSW technique, to provide at once a comprehensive picture of the ferroelectric polarization at (and near) the surface. This type of study can be applied to other interesting oxides and can be extended to a broader class of other technologically relevant materials, such as multiferroics.\textsuperscript{\cite{spaldin_advances_2019}} Moreover, in the context of catalytic reactions at ferroelectric surfaces the determination and control of surface polarization and the interplay with adsorbates is crucial. To this end, \textit{in operando} XSW investigations can guide material engineering towards more efficient catalysts.\textsuperscript{\cite{kakekhani_ferroelectric-based_2015, kakekhani_polarization-driven_2016}} Importantly, the few pm structural accuracy of XSW provides a rigorous test for bechmarking different theoretical models and thereby improving their predictive power.\textsuperscript{\cite{Gattinoni20, Deleuze20}} Finally, we anticipate that the XSW technique can be employed to investigate the dynamics of ferroelectric polarization switching, specifically to simultaneously track structural and electronic changes of atoms in real time and corresponding measurements at X-ray free-electron laser facilities are in preparation.

\end{justify}


\medskip
\textbf{Acknowledgements} \par
\begin{justify}
We acknowledge the Diamond Light Source Ltd. for beamtime at the I09 beamline under the SI27468-1 proposal and thank the staff for their assistance during our experiments. D.P acknowledges funding from ‘la Caixa’ Foundation fellowship (ID 100010434). I.S. and D.P. acknowledge financial support from the Spanish Ministerio de Ciencia e Innovacion (MICINN), grant No. PID2019-109931GB-I00. The ICN2 is funded by the CERCA programme / Generalitat de Catalunya and by the Severo Ochoa Centres of Excellence Programme, funded by the Spanish Research Agency (AEI, CEX2021-001214-S).  
\end{justify}

\medskip
\textbf{Conflict of Interest} \par
The authors declare no conflict of interest.

\medskip
\textbf{Data Availability Statement} \par
The data that support the findings of this study are available from the corresponding author upon reasonable request.

\begin{justify}
\bibliographystyle{MSP}
\bibliography{references}
\end{justify}

\newpage

\begin{figure}[h]
\centering
\includegraphics[width=0.9\textwidth]{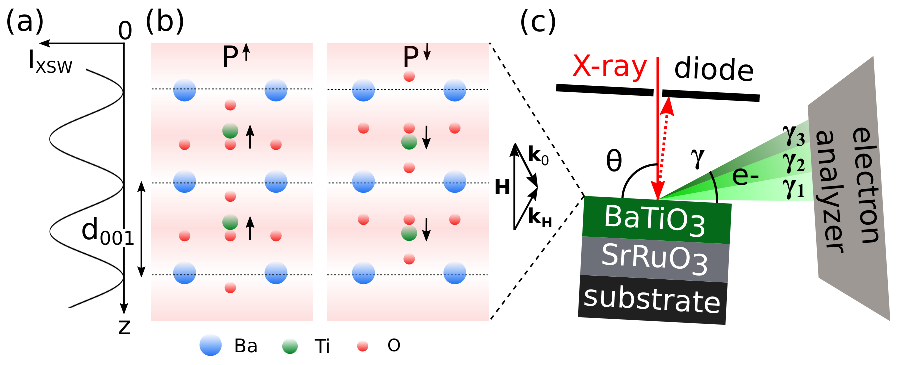}
\begin{justify}
\caption{(a) XSW intensity of BTO (001) Bragg reflection and $z$ axis orientation, with $z = 0$ at the sample surface. (b) Side view of the top two BTO unit cells with ferroelectric polarization $\mathrm{P}^\uparrow$ and $\mathrm{P}^\downarrow$, and Bragg spacing $d_{001}$. (c) Sketch of the experimental setup (top view) used at the beamline I09 of the Diamond Light Source, including sample, electron analyzer, and photodiode. The photodiode was located \SI{10}{\mm} away from the sample and was equipped with an Al mask in front to minimize the fluorescence background. The Bragg angle $\theta$ and the photoelectron exit angle $\gamma$ are shown, together with photoelectron exit angle ranges $\gamma_1$, $\gamma_2$ and $\gamma_3$, incident $\boldsymbol{k}_0$ and Bragg-diffracted $\boldsymbol{k}_{\boldsymbol{H}} = \boldsymbol{k}_0 + \boldsymbol{H}$ X-ray wavevectors.}\label{fig:setup}
\end{justify}
\centering
\end{figure}

\begin{figure}[b]
\centering
\includegraphics[width=1\textwidth]{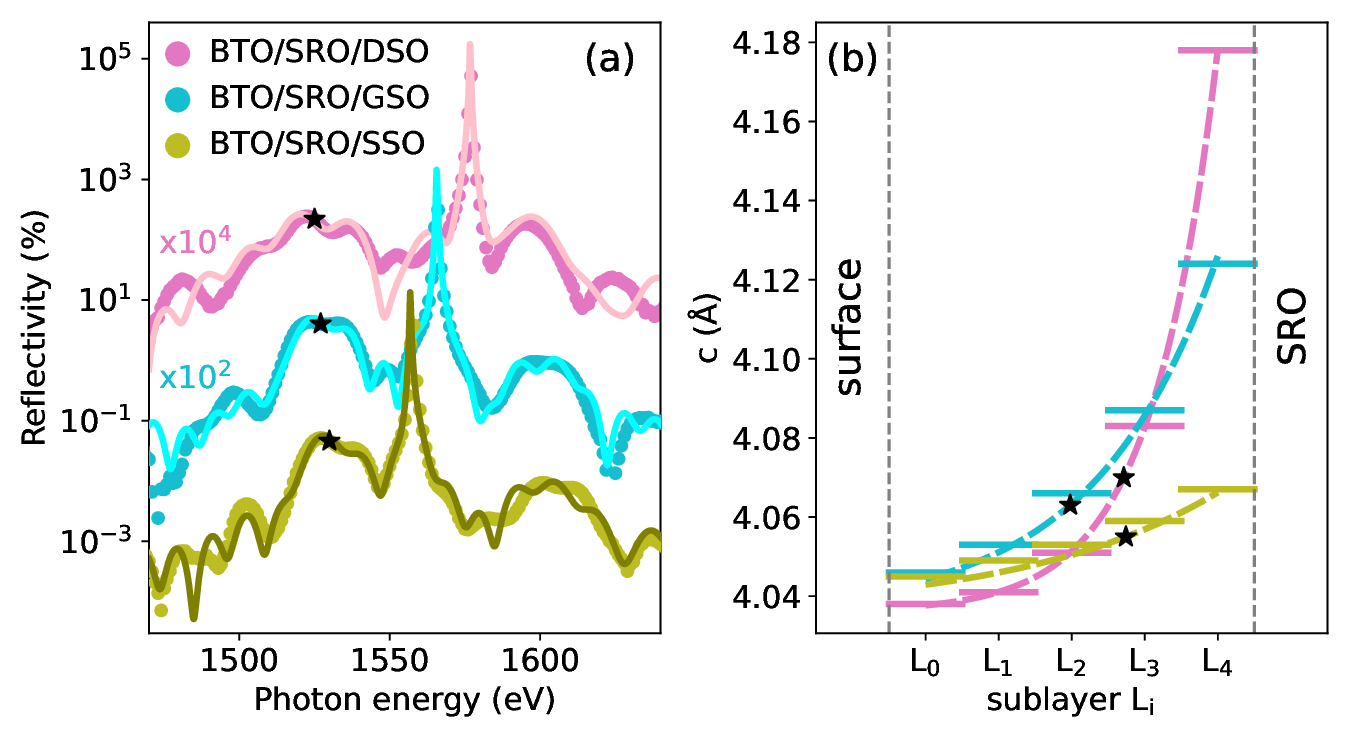}
\begin{justify}
\caption{(a) (001) Bragg reflectivity $R_0(E_\nu)$ (points) and corresponding fit curves (solid lines) of samples BTO/SRO/DSO (pink), BTO/SRO/GSO (cyan) and BTO/SRO/SSO (green). (b) BTO out-of-plane lattice parameter $c_i$ (solid lines) in sublayers $L_i$ and $c(z)$ (dotted lines) based on Equation \eqref{eq:c-gradient}. The average BTO out-of-plane lattice parameters $\overline{c}_\mathrm{BTO}$ are marked by black stars.}\label{fig:xrd-c}
\end{justify}
\centering
\end{figure}

\begin{figure}[b]
\centering
\includegraphics[width=1\textwidth]{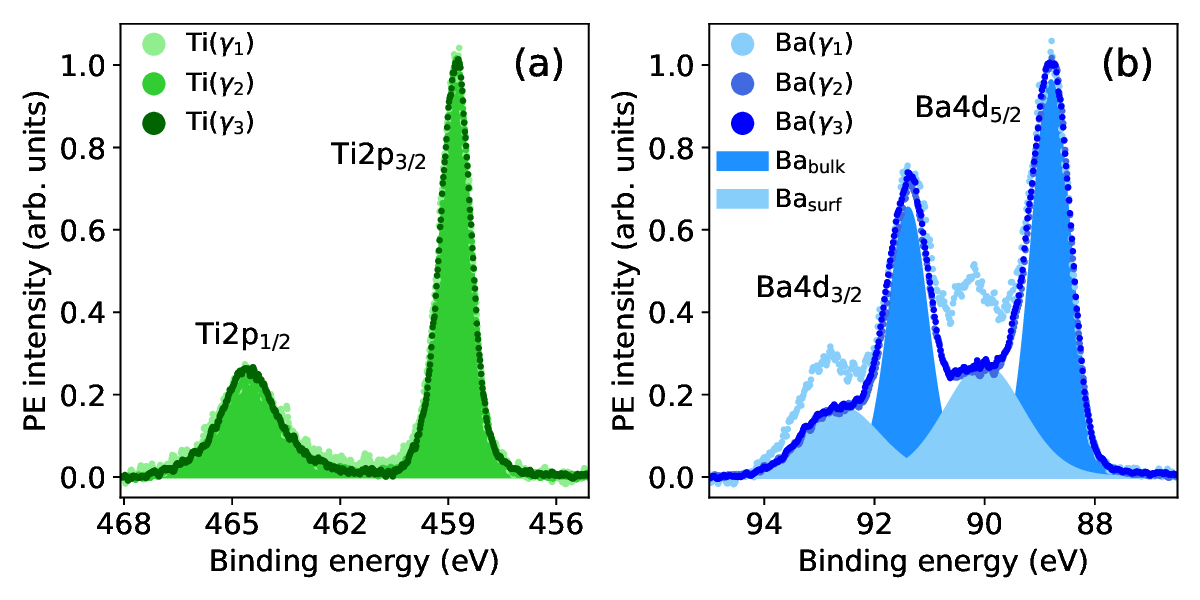}
\begin{justify}
\caption{PE spectra of Ti 2p (a) and Ba 4d (b) core levels measured with $E_\nu = $ \SI{1420}{\eV} at the exit angle ranges $\gamma_1$, $\gamma_2$, and $\gamma_3$, on the BTO/SRO/DSO sample. Each spectrum is normalized to the respective PE intensity maximum. Shaded component areas refer to spectra measured at the exit angle range $\gamma_1$. The $\mathrm{Ba_{surf}}$ [$\mathrm{Ba_{bulk}}$] component refers to Ba atoms at [below] the top BaO atomic plane. Similar PE spectra measured on the BTO/SRO/GSO and BTO/SRO/SSO samples are reported in Figure \ref{SM:fig:xps_sam4_sam5} (Supporting Information).}\label{fig:XPS-Ba-Ti}
\end{justify}
\centering
\end{figure}

\begin{figure*}[b]
\centering
\includegraphics[width=1\textwidth]{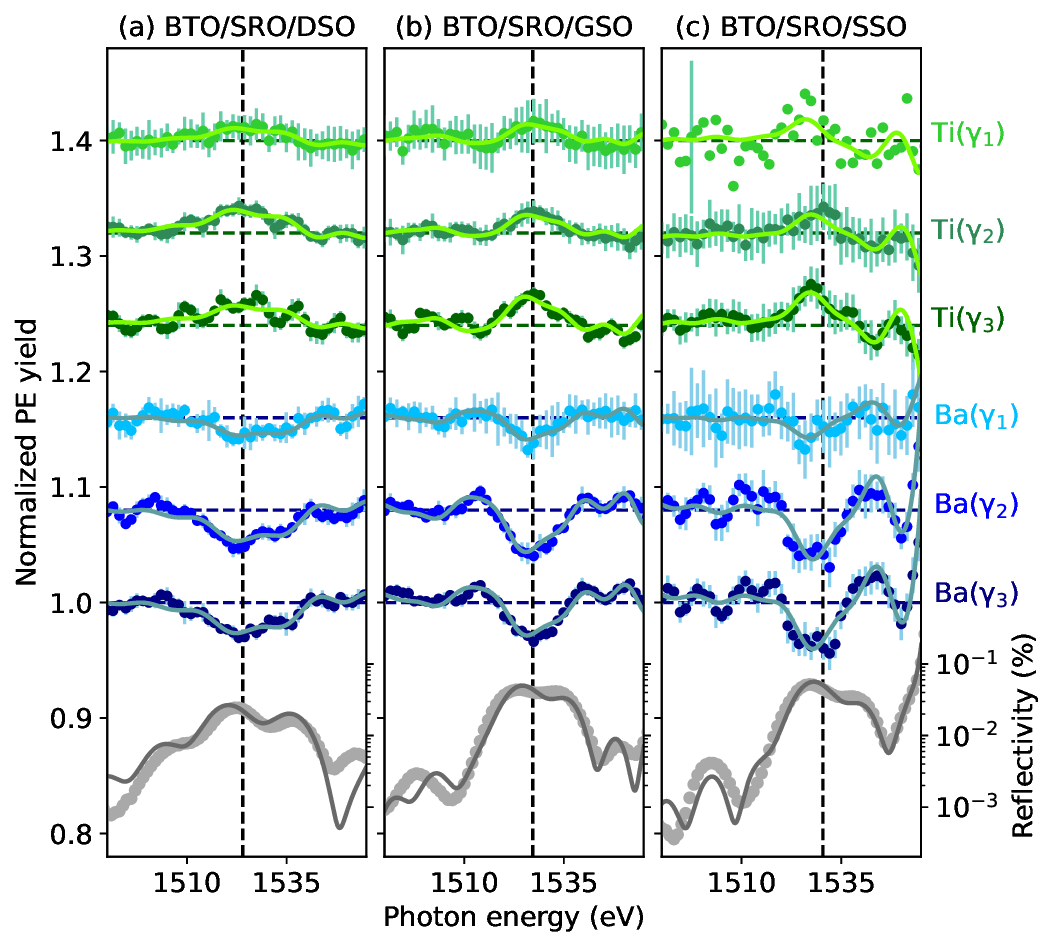}
\begin{justify}
\caption{Ti and Ba PE yield data (green and blue points) measured with $E_\nu =$ \SI{1420}{\eV} at the exit angle ranges $\gamma_1$, $\gamma_2$, and $\gamma_3$ on BTO/SRO/DSO (a), BTO/SRO/GSO (b), and BTO/SRO/SSO (c), and corresponding fit curves (solid lines). Reflectivity $R_0(E\nu)$ data (gray points) and fit curves (solid gray lines) around the (001) BTO Bragg energies $E_B =$ \SI{1524}{\eV} (a), \SI{1527.2}{\eV} (b), \SI{1530.4}{\eV} (c) (marked by vertical dashed lines). For clarity, $\kappa^{\mathrm{Ti}}_{\gamma_1}(E_\nu)$ of BTO/SRO/SSO is shown with only one error bar, which corresponds to the average error bar of all $\kappa^{\mathrm{Ti}}_{\gamma_1}(E_\nu)$ data points. All PE yield curves are normalized (Section  \ref{SM:sec:yield-normalization}, Supporting Information) and, for clarity, the curves above $\mathrm{Ba}(\gamma_3)$ are vertically shifted by $0.08$ from the one below.}\label{fig:XSW-data}
\end{justify}
\centering
\end{figure*}

\begin{figure}[t]
\centering
\includegraphics[width=1\textwidth]{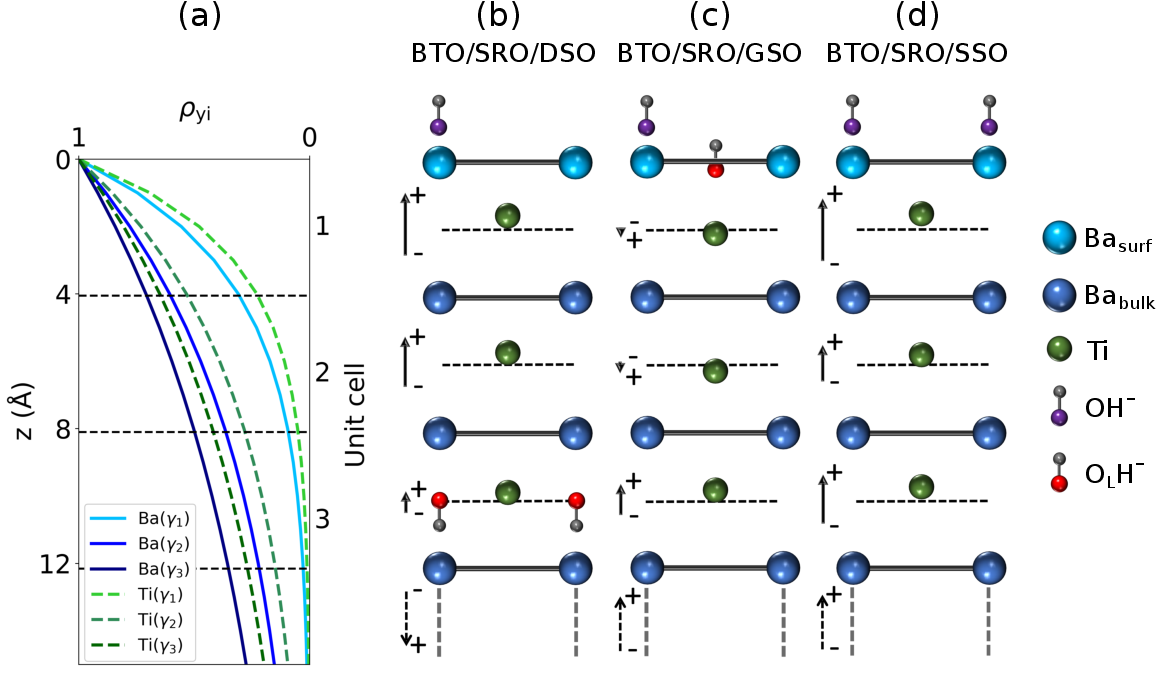}
\begin{justify}
\caption{(a) Probability yield functions $\rho_\mathrm{yi}(z)$ of Ba 4d and Ti 2p photoelectrons at $E_B = $ \SI{1525}{\eV} integrated over the three exit angle ranges $\gamma_1$, $\gamma_2$ and $\gamma_3$. Sketch of Ba and Ti atoms in the top three unit cells of BTO/SRO/DSO (b), BTO/SRO/GSO (c), and BTO/SRO/SSO (d). For a better visualization, Ti atomic displacements $\Delta z^{\mathrm{Ti}}_\gamma$ are twice larger than values in Table \ref{tab:Pc-Fc}. The length of polarization vectors (solid arrows) is proportional to the corresponding $\Delta z^{\mathrm{Ti}}_\gamma$. The direction of the average ferroelectric polarization in BTO films measured by PFM is marked by dashed arrows below the third unit cell. For clarity, only $\mathrm{OH^-}$ and $\mathrm{O_LH^-}$ species are sketched, while $\mathrm{O_2^-}$ species and $\mathrm{O_L}$ atoms are omitted.}\label{fig:sketch}
\end{justify}
\centering
\end{figure}

\appendix

\clearpage

\begin{justify}

\title{Supporting Information}

\section{Sample growth and characterization}
\subsection{Sample growth}
\label{SM:sec:LPD}

Epitaxial bilayers BTO/SRO are grown on DSO, GSO, SSO substrates using pulsed laser deposition. The ceramic targets of SRO and BTO were 8 cm away from the substrates and ablated using a KrF excimer laser ($\lambda$ = \SI{248}{\nm}, fluence \SI{5.4}{\joule\per\square\cm}, \SI{2}{\hertz} repetition rate). The deposition of SRO and BTO layer is conducted in O$_2$ atmosphere with pressure pO$_2$ = 100 mTorr and deposition temperature of \SI{908}{\kelvin} and \SI{973}{\kelvin}, respectively. Sample cooling with the rate of \SI{3}{\kelvin\per\minute} is conducted in the environment of saturated O$_2$ (pO$_2$ = $10^4$ mTorr) to prevent the formation of oxygen vacancies.

\subsection{Grazing X-ray reflectivity}
\label{SM:sec:XRR}

\begin{figure*}[b!]
\centering
\includegraphics[width=0.6\textwidth]{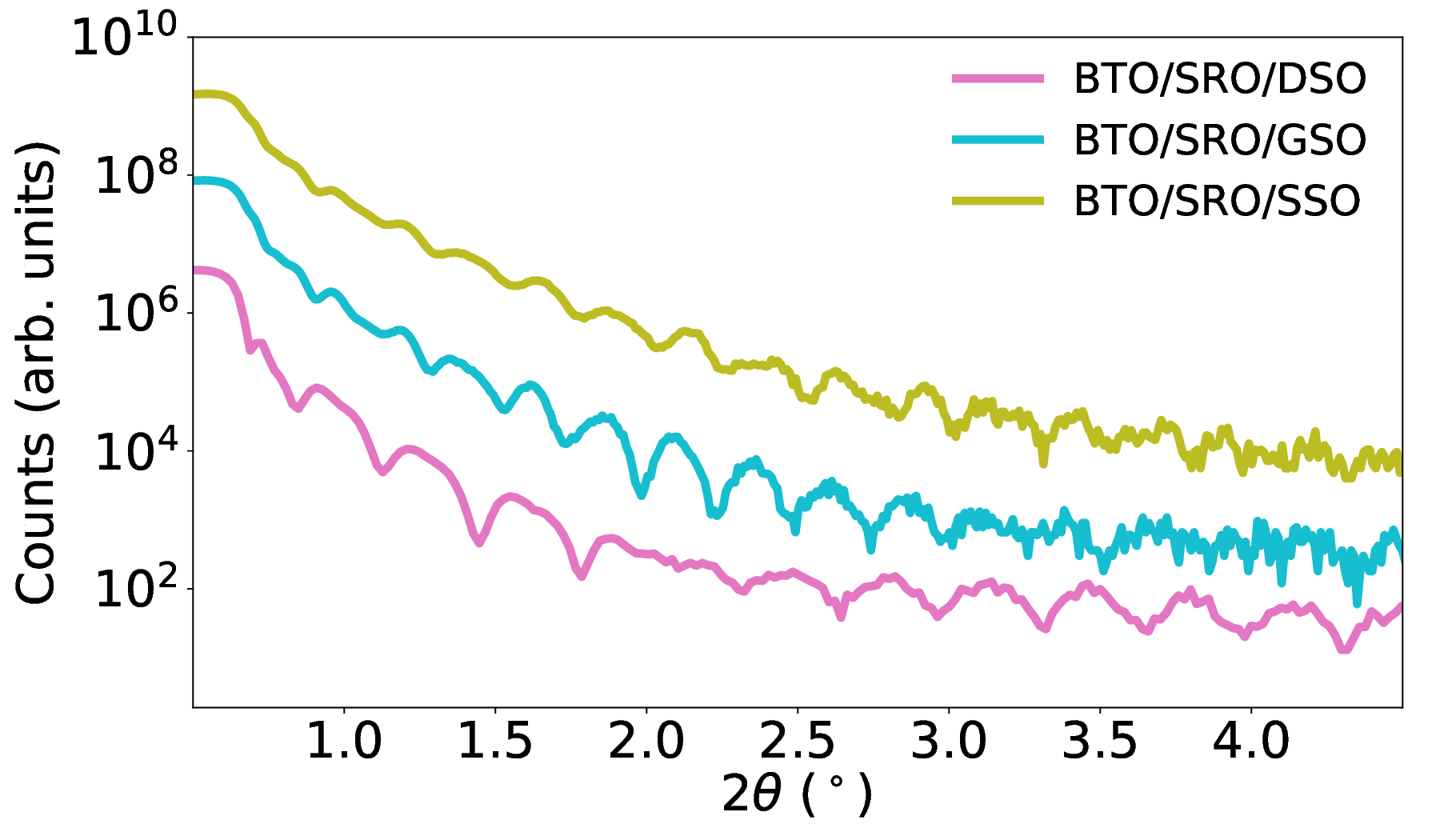}
\caption{Grazing X-ray reflectivity data of as-grown samples used to determine the thickness of the BTO and SRO thin films (Table \ref{tab:thickness-strain}).}
\centering
\label{SM:fig:XRR}
\end{figure*}

Grazing X-ray reflectivity data of as-grown samples, measured by a PANalytical X'Pert Pro diffractometer, are shown in Figure \ref{SM:fig:XRR}. The measured reflectivity $R_g(q)$ can be expressed as 
\begin{equation}
  R_g(q) = R_F(q)\Big|\frac{1}{\rho_s}\int_{-\infty}^{\infty}\frac{d\rho_e(z)}{dz}\exp{(-\mathrm{i}qz)dz}\Big|^2,
\label{sm:eq:XRR}
\end{equation}
where $R_F$ and $\rho_s$ are the Fresnel reflectivity and the electron density of the substrate, respectively \cite{vignaud_ordering_1998}. In Equation \eqref{sm:eq:XRR} $R_g$ is expressed as a function of the wavevector transfer $q = (4\pi \sin{\theta})/\lambda$ , where $\theta$ is the incident angle of X-ray and $\lambda = $ \SI{1.54}{\angstrom} is the wavelength of Cu K$\alpha$ incident radiation. In practice, thin film thicknesses are determined as follows. First, $R_F$ is calculated using the Parratt formalism \cite{Parratt54}. Second, the Fourier inversion of $R_g/R_F$ provides the autocorrelation of the derivative of the electron density $\rho'_e$ as a function of $z$. This function displays peaks in correspondence of the interfaces, where $\rho'_e$ is largest, thereby providing the thicknesses of layers above the substrate.

\subsection{X-ray reciprocal lattice map}
\label{SM:sec:RSM}

\begin{figure*}[t!]
\centering
\includegraphics[width=\textwidth]{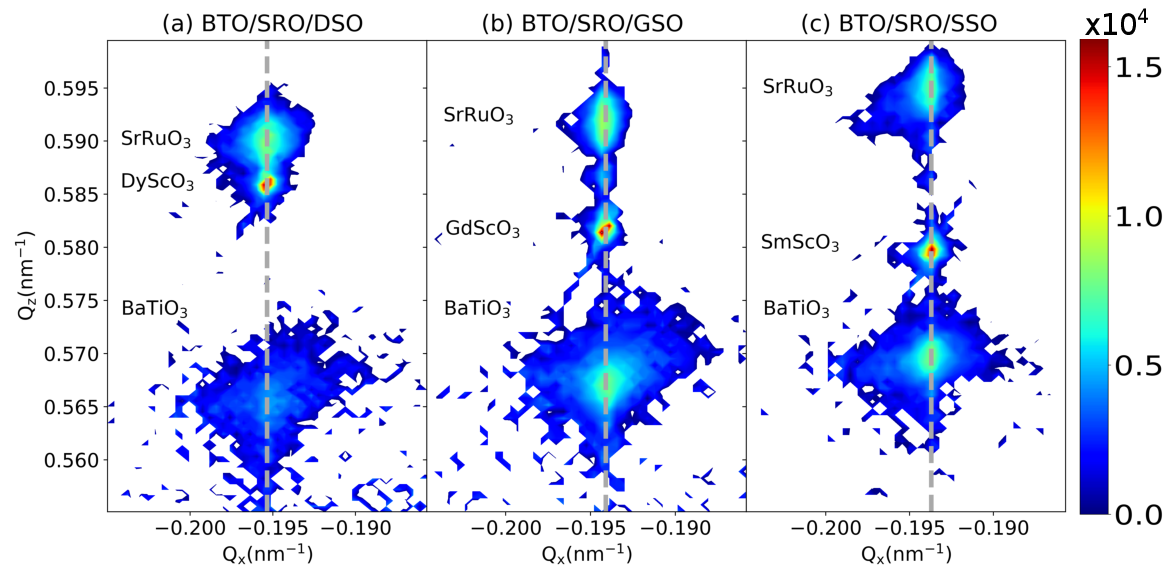}
\caption{Reciprocal space maps around (-103) substrate Bragg peak of the BTO/SRO/DSO (a), BTO/SRO/GSO (b) and BTO/SRO/SSO (c) samples. The gray vertical dashed lines indicate the reciprocal lattice parameter $Q_\mathrm{x}$ shared by the substrate, the BTO, and SRO thin films in each sample.}
\centering
\label{SM:fig:RSM}
\end{figure*}

Figure \ref{SM:fig:RSM} shows X-ray reciprocal lattice maps of the three samples around (-103) substrate Bragg peak, measured by a PANalytical X'Pert Pro diffractometer. The diffraction peaks of the substrates have very narrow intensity distribution, while the intensity distribution of BTO and SRO layers are weaker and broader. Reciprocal lattice parameters $Q_\mathrm{x}$ and $Q_\mathrm{z}$ of intensity peaks are related to the real space in-plane and out-of-plane lattice parameters, $a$ and $c$, by the following relations: $a = -\lambda/(2Q_\mathrm{x})$ and $c = (3\lambda)/(2Q_\mathrm{z})$ \cite{Mario05}. The values $Q_\mathrm{x}$ and $Q_\mathrm{z}$ of each diffraction peak are obtained by fitting the intensity distribution with a pseudo-Voigt function \cite{Young82}.

\subsection{Piezoresponse force microscopy}
\label{SM:sec:pfm}

\begin{figure*}[t!]
\centering
\includegraphics[width=1\textwidth]{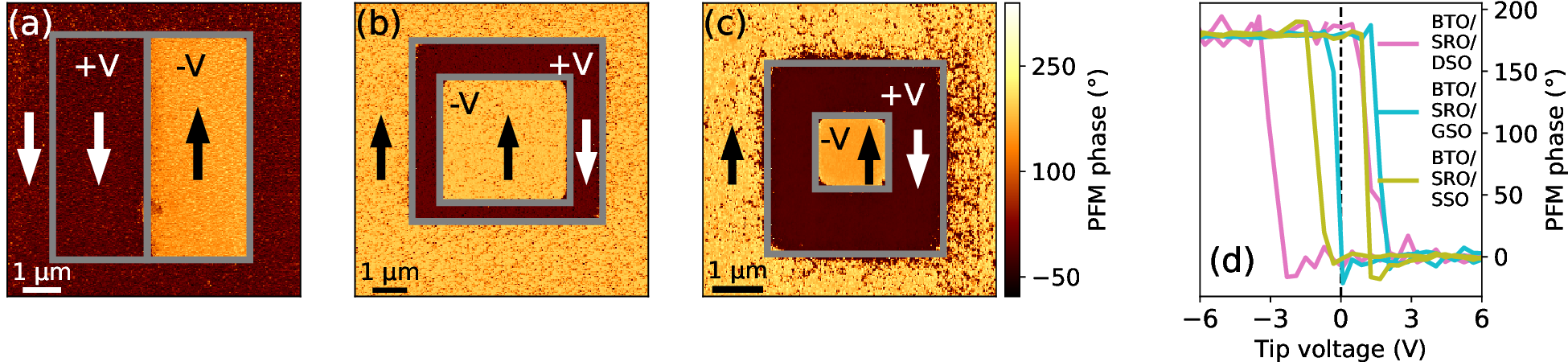}
\caption{PFM phase image of the BTO/SRO/DSO (a), BTO/SRO/GSO (b), and BTO/SRO/SSO (c) samples. The sign of the applied tip voltage ($\pm \mathrm{V}$) within the gray boxes and the resulting average polarization direction $\mathrm{P}^\uparrow$ or $\mathrm{P}^\downarrow$ in the probed areas are marked on each panel. The PFM phase beyond the gray boxes indicates the average polarization direction of as-grown samples. (d) The corresponding hysteresis loops.}
\centering
\label{SM:fig:pfm}
\end{figure*}

Piezoresponse force microscopy in Dual AC Resonance Tracking (DART) mode \cite{rodriguez_dual-frequency_2007} was used to probe the polarization of as-grown samples and to prove that ferroelectric polarization can be switched by the application of positive or negative voltage between the PFM tip and the SRO electrode. Figure \ref{SM:fig:pfm}a-c show the PFM phase image of each sample after the application of a voltage to switch the polarization inside the marked gray boxes. A positive [negative] voltage of sufficient amplitude forces the polarization to be $\mathrm{P}^\downarrow$ [$\mathrm{P}^\uparrow$], with the corresponding PFM phase $0\degree$ [$180\degree$]. The PFM phase outside the gray boxes indicates the polarization of the as-grown sample. Therefore, Figure \ref{SM:fig:pfm}a shows an average downward polarization $\mathrm{P}^\downarrow$ in the BTO/SRO/DSO sample, while an average upward polarization $\mathrm{P}^\uparrow$ in the BTO/SRO/GSO and BTO/SRO/SSO samples. Note also that PFM phase images show no indication of multiple domains in any of the as-grown sample. Furthermore, switching spectroscopy PFM (SS-PFM) was employed to measure hysteresis loops on each sample. From these data, reported in Figure \ref{SM:fig:pfm}d, we measure in the BTO/SRO/DSO sample a coercive voltage of $V_c =$ \SI{1.1}{\volt} with a negative bias $V_b =$ \SI{-0.9}{\volt}. In contrast, the BTO/SRO/GSO sample shows $V_c =$ \SI{1.6}{\volt} with a positive bias $V_b =$ \SI{0.7}{\volt}. Finally, the BTO/SRO/SSO sample has an unbiased hysteresis loop with a coercive voltage $V_c =$ \SI{1.1}{\volt}.

\section{Data analysis}

\subsection{Average out-of-plane lattice parameter $\overline{c}$}
\label{SM:sec:c_average}

BTO and SRO average out-of-plane parameters $\overline{c}$ are calculated from the corresponding (001) Bragg peaks using the Bragg condition $\overline{c} = (\SI{12400}{\eV\angstrom})/(2\overline{E}_\nu\sin\theta_B)$. $\overline{E}_\nu$ is the average of energy values $E_\nu$ around the Bragg peaks, weighted with reflectivity $R_0(E_\nu)$. The energy ranges for calculating $\overline{E}_\nu$ of BTO and SRO are respectively $1510$ \SI{}{\eV} - $1547.5$ \SI{}{\eV} and $1587$ \SI{}{\eV} - $1624$ \SI{}{\eV}, where the $R_0(E_\nu)$ has finite values.

\subsection{Reflectivity fit results}
\label{SM:sec:reflfit}

Table \ref{SM:tab:reflfit_W} and Table \ref{SM:tab:reflfit_eps_delta} report the results of reflectivity fit shown in Figure \ref{fig:xrd-c}a.

\begin{table*}[h!]
\caption{\label{SM:tab:reflfit_W} BTO and SRO Debye-Waller factors $\mathrm{e}^{-W_i}$ of sublayers $L_i$ ($i=0, ...,4$) resulting from the fits of (001) Bragg reflection data in the three samples under study (Figure \ref{fig:xrd-c}).}
\begin{tabular}{cccccccc}
 \hline
 Layer & sublayer & fit parameters & BTO/SRO/DSO & BTO/SRO/GSO & BTO/SRO/SSO \\ 
 \hline
 & $L_0$ &$\mathrm{e}^{-W_0}$ &  1 & 0.9  & 1 \\
 & $L_1$ &$\mathrm{e}^{-W_1}$ &  1 & 1  & 1 \\
 BTO & $L_2$ &$\mathrm{e}^{-W_2}$ &  1 & 1  & 0.8 \\
 & $L_3$ &$\mathrm{e}^{-W_3}$ &  1 & 1  & 1 \\
 & $L_4$ &$\mathrm{e}^{-W_4}$ &  1 & 0.1  & 0.5 \\
\hline
 & $L_0$ &$\mathrm{e}^{-W_0}$ &  0.4 & 0.7  & 0.9 \\
 & $L_1$ &$\mathrm{e}^{-W_1}$ &  0.9 & 0.6  & 1 \\ 
 SRO & $L_2$ &$\mathrm{e}^{-W_2}$ &  1 & 0.9  & 0.6 \\ 
 & $L_3$ &$\mathrm{e}^{-W_3}$ &  0.4 & 1  & 0.5 \\ 
 & $L_4$ &$\mathrm{e}^{-W_4}$ &  0.04 & 0  & 0.8 \\
 \hline
\end{tabular}
\end{table*}

\begin{table*}[h!]
\caption{\label{SM:tab:reflfit_eps_delta} BTO and SRO interface strain $\epsilon_\mathrm{int}$ and penetration depth of strain $\delta$ resulting from the fits of (001) Bragg reflection data in the three samples under study (Figure \ref{fig:xrd-c}).}
\begin{tabular}{ccccccc}
 \hline
 Layer &  fit parameters & BTO/SRO/DSO & BTO/SRO/GSO & BTO/SRO/SSO \\ 
 \hline
BTO &$\epsilon_\mathrm{int}$ & 0.06 & 0.03 & 0.01 \\
 &$\delta$ & 36 & 136 & 233 \\ 
\hline
SRO &$\epsilon_\mathrm{int}$ & -0.01 & -0.05 & -0.04 \\
 &$\delta$ & 589 & 56 & 71 \\ 
 \hline
\end{tabular}
\end{table*}

\subsection{SRO structural properties}
\label{SM:sec:SRO_structure}

\begin{figure*}[b!]
\centering
\includegraphics[width=0.6\textwidth]{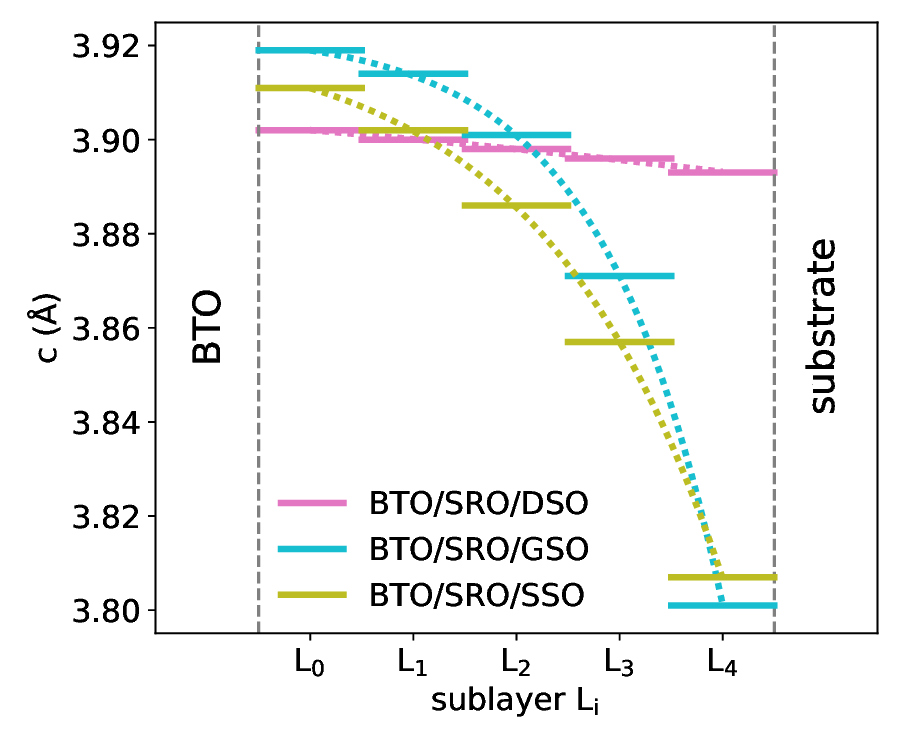}
\caption{SRO out-of-plane lattice parameters $c_i$ (solid lines) in sublayers $L_i$ and $c(z)$ (dotted lines) according to Equation \eqref{eq:c-gradient}.} 
\centering
\label{SM:fig:c_SRO}
\end{figure*}

The in-plane strain applied by a substrate to the BTO thin film is calculated as $\epsilon^{a}_\mathrm{SRO} = \left(a_\mathrm{SRO} - a_\mathrm{b,SRO} \right) / a_\mathrm{b,SRO}$, by comparing the measured in-plane lattice parameter of the thin film $a_\mathrm{SRO}$ (Section \ref{SM:sec:RSM}) with the respective bulk value of the pseudocubic cell with parameter $a_\mathrm{b,SRO} \approx c_\mathrm{b,SRO} =$ \SI{3.923}{\angstrom} \cite{Jia02}. As a result, the DSO, GSO, and SSO substrates impose an in-plane tensile strain on the SRO films of $0.51 \%$, $1.12 \%$, and $1.38 \%$, respectively. This leads to a decreasing average out-of-plane lattice parameter in SRO following the same substrate order: $\overline{c}_\mathrm{SRO} =$  \SI{3.885 \pm 0.019}{\angstrom}, \SI{3.876 \pm 0.024}{\angstrom}, \SI{3.866 \pm 0.023}{\angstrom}. These measured $\overline{c}$ parameters correspond to an average out-of-plane tensile strain $\overline{\epsilon}^c_\mathrm{SRO} = \left( \overline{c}_\mathrm{SRO} - c_{b,\mathrm{SRO}} \right) / c_{b,\mathrm{SRO}}$ of $-0.96 \%$, $-1.20 \%$, and $-1.45 \%$, respectively. Figure \ref{SM:fig:c_SRO} shows the exponential distribution of SRO out-plane lattice parameters $c_i$ (Equation 4 in the article) in different sublayers $L_i$, resulting from the (001) Bragg reflections fits shown in Figure \ref{fig:xrd-c}a.

\subsection{Ti 2p, Ba 4d and C 1s XPS}
\label{SM:sec:Ti-Ba-C-XPS}

\begin{figure*}[b!]
\centering
\includegraphics[width=1\textwidth]{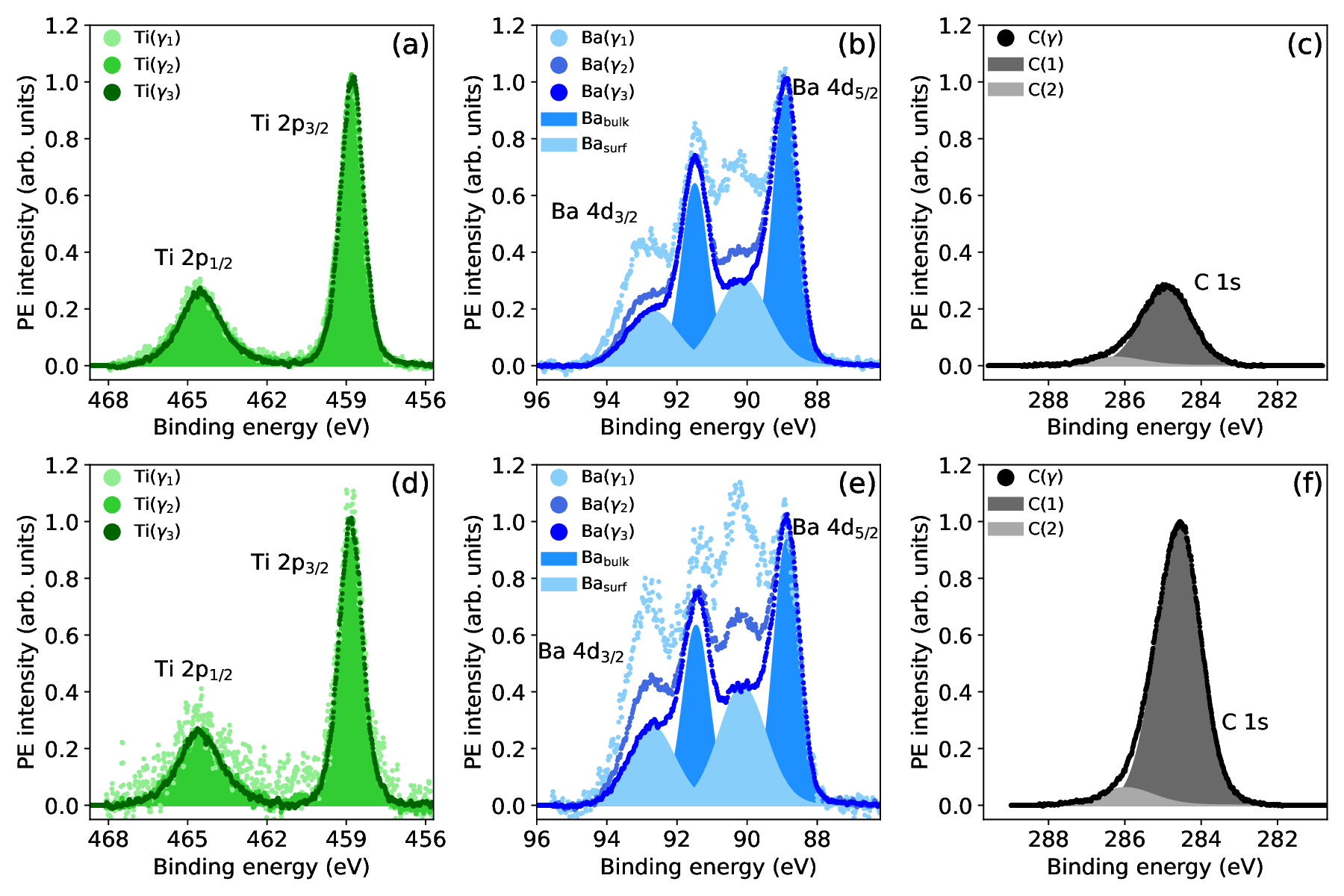}
\caption{Ti 2p, Ba 4d and C 1s PE spectra of the BTO/SRO/GSO sample (a, b, c) and the BTO/SRO/SSO sample (d, e, f), respectively. Ti 2p and Ba 4d spectra are shown at the exit angle ranges $\gamma_1$, $\gamma_2$, and $\gamma_3$, while panels c and f show the sum of C 1s spectra over all measured exit angle ranges. Note that both C 1s spectra are normalized to the peak intensity of C 1s spectrum in panel (f) to underscore the different content of carbon species in the two samples.}
\centering
\label{SM:fig:xps_sam4_sam5}
\end{figure*}

Figure \ref{SM:fig:xps_sam4_sam5} shows typical PE spectra Ti 2p, Ba 4d, and C 1s core levels of the BTO/SRO/GSO and BTO/SRO/SSO samples. Ti and Ba PE spectra show similar features as the spectra in Figure \ref{fig:XPS-Ba-Ti} discussed in Section \ref{sec:results-xps}. The larger noise of PE spectra in Figure \ref{SM:fig:xps_sam4_sam5}d-e result from the lower PE intensity due to the larger amount of C and O species on the BTO/SRO/SSO surface. In fact, the corresponding C 1s spectrum (Figure \ref{SM:fig:xps_sam4_sam5}f) has an area approximately three times larger than of the one of the BTO/SRO/GSO sample (Figure \ref{SM:fig:xps_sam4_sam5}c). In particular, component $\mathrm{C(1)}$ at \SI{284.6}{\eV} is assigned to adventitious carbon, resulting from hydrocarbons adsorbed on the sample surface, upon exposure to ambient environment, while component $\mathrm{C(2)}$ with a BE shift of \SI{1.4}{\eV} refers to $\mathrm{CO}$ species \cite{Irena21}. The C1s spectrum of the BTO/SRO/DSO sample (not shown) is similar to the one of the BTO/SRO/GSO sample (Figure \ref{SM:fig:xps_sam4_sam5}c).

\subsection{O 1s XPS}
\label{SM:sec:results-O1s-xps}

O 1s XPS spectra measured at different exit angle range $\gamma$ (Figure \ref{SM:fig:xps-O1s}a-c) provide a detailed picture of oxygen-related species at the BTO surface. Let us focus first on the most bulk sensitive $\mathrm{O}(\gamma_3)$ spectra. Here, the most prominent peak $\mathrm{O(1)}$ at \SI{529.8}{\eV} refers to O atoms in BTO lattice ($\mathrm{O_L}$). The other O components result from H$_2$O dissociation ($\mathrm{O(2)}$ and $\mathrm{O(4)}$) and $\mathrm{CO_x}$ species ($\mathrm{O(3)}$). In general, adsorption of $\mathrm{H_2O}$ on BTO may lead to molecular physisorption with $\mathrm{\Delta BE}\approx$ \SI{3.9}{\eV} \cite{domingo_water_2019,Irena21}. O 1s spectra in Figure \ref{SM:fig:xps-O1s}a-c show evidence of molecular physisorbed water only at the exit angle ranges $\gamma_1$ and $\gamma_2$ and with minor contributions to the total spectral area smaller than $3\%$. This suggests that the surface favors water dissociation into OH$^-$ and H$^+$, which is typical of BaO-terminated surfaces \cite{geneste_adsorption_2009,shin_atomistic_2009} but is also present on $\mathrm{TiO_2}$-terminated surfaces \cite{domingo_water_2019}. While OH$^-$ chemisorbs on top of cations (Ba or Ti) or at O vacancies, H$^+$ binds to $\mathrm{O_L}$ at the BTO surface or diffuses inside the film to form $\mathrm{O_LH^-}$ \cite{Wang12, lee_imprint_2016, park_effect_2000}, assigned to component $\mathrm{O(2)}$. The latter has a smaller binding energy shift $\mathrm{\Delta BE} =$ \SI{1}{\eV} \cite{chun_surface_2022}, given the chemical environment much closer to $\mathrm{O_L}$, as compared to chemisorbed OH$^-$. In contrast, OH$^-$ is assigned to component $\mathrm{O(4)}$ with a larger binding energy shift $\mathrm{\Delta BE} =$ \SI{2.7}{\eV} \cite{wang_chemistry_2012, lee_imprint_2016}. In addition, other species, resulting from $\mathrm{O_2}$ adsorption such as peroxo complexes (e.g. BaO$_2$, Ti-O-O-Ti, Ti=O$_{2}^{-}$), can contribute to component $\mathrm{O(4)}$ \cite{domingo_water_2019}. In this case, oxidation of BaO-terminated surfaces or Ti-OH leads to the presence of O$_{2}^{-}$ at the surface. Finally, component $\mathrm{O(3)}$ at $\mathrm{\Delta BE} =$ \SI{1.8}{\eV} is assigned to CO$_\mathrm{x}$ species, such as carbonates $\mathrm{CO_3^{2-}}$, $\mathrm{C=O}$ bonds, ester (C-(C=O)-OR) or carboxylic acid (C-(C=O)-OH) compounds \cite{landoulsi_organic_2016, Irena21}. In the BTO/SRO/DSO, BTO/SRO/GSO, and BTO/SRO/SSO samples, the contribution of component $\mathrm{O(3)}$ to the $\gamma$-integrated spectral area is $\approx 0 \%$, $3 \%$, and $6 \%$, respectively. Following the same substrate order, component $\mathrm{O(2)}$ [$\mathrm{O(4)}$] represents $15 \%$ [$32 \%$], $19 \%$ [$30 \%$], and $13 \%$ [$53 \%$] of the $\gamma$-integrated spectral area.

\begin{figure}[b!]
\centering
\includegraphics[width=1\textwidth]{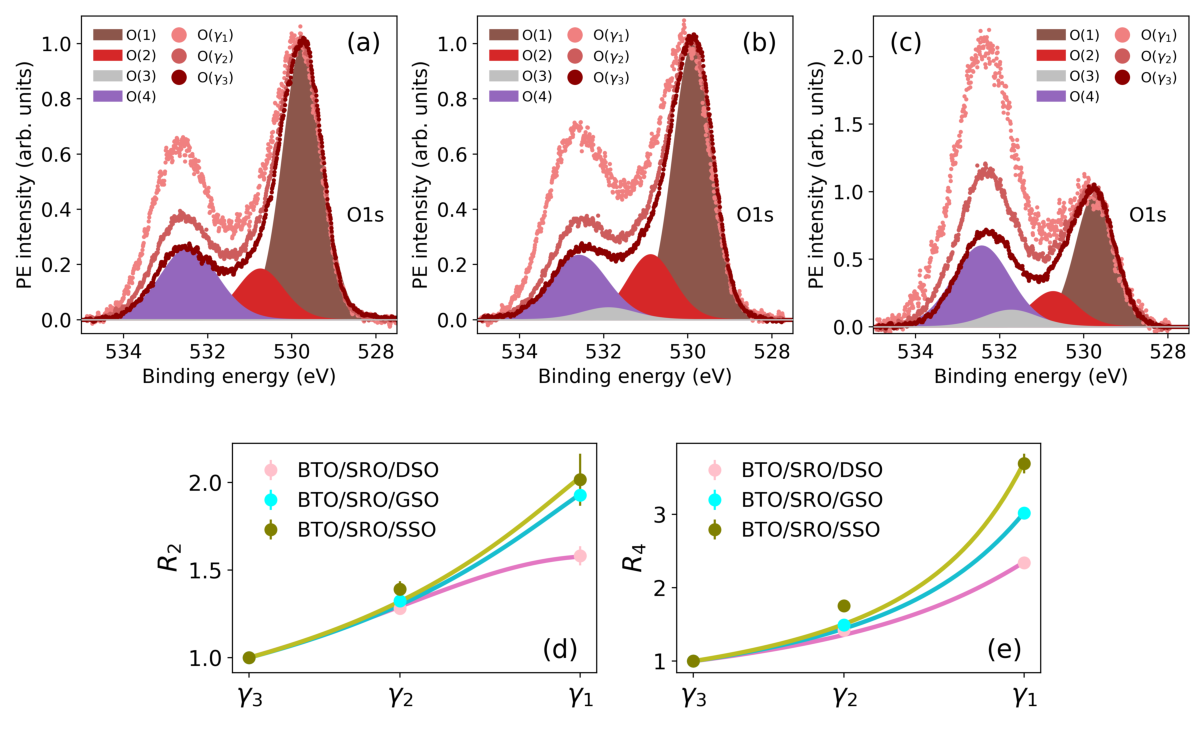}
\caption{O 1s PE spectra $\mathrm{O}(\gamma_1)$, $\mathrm{O}(\gamma_2)$ and $\mathrm{O}(\gamma_3)$ integrated over the respective exit angle ranges in the BTO/SRO/DSO (a), BTO/SRO/GSO (b), and BTO/SRO/SSO (c) samples. Each spectrum is normalized to the respective maximum PE intensity. Shaded component areas $\mathrm{O(1)}$, $\mathrm{O(2)}$, $\mathrm{O(3)}$, and $\mathrm{O(4)}$ refer to $\mathrm{O}(\gamma_3)$ spectra. Components $\mathrm{O(1)}$, $\mathrm{O(2)}$, $\mathrm{O(3)}$, and $\mathrm{O(4)}$ originate from O atoms in the lattice ($\mathrm{O_L}$), $\mathrm{O_LH^-}$ species, $\mathrm{CO_x}$ species, and $\mathrm{OH^-}$ and/or $\mathrm{O_2^-}$ species, respectively. The ratio of component $\mathrm{O(2)}$ [$\mathrm{O(4)}$] over component $\mathrm{O(1)}$ as a function of exit angle range $\gamma$ is displayed in panel (d) [(e)] for the three samples under study.}
\centering  
\label{SM:fig:xps-O1s}
\end{figure}

To gain information on the depth distribution of the different oxygen species, a quantitative analysis of $\mathrm{O(2)}$ and $\mathrm{O(4)}$ relative spectral area as a function of exit angle range $\gamma$ was performed. The ratio of spectral areas between component $\mathrm{O(2)}$ [$\mathrm{O(4)}$] and $\mathrm{O(1)}$ measured at exit angle range $\gamma_j$ ($j = 1, 2, 3$) is calculated as $ R_2 = A_{\mathrm{O(2)}}^{\gamma_j} / A_{\mathrm{O(1)}}^{\gamma_j}$ [$ R_4 = A_{\mathrm{O(4)}}^{\gamma_j} / A_{\mathrm{O(1)}}^{\gamma_j}$]. Figure \ref{SM:fig:xps-O1s}d,e shows $R_2$ and $R_4$ at the different $\gamma_j$ together with the corresponding fit functions. The model employed to fit these data \cite{fadley_angle-resolved_1984} assumes that adsorbates form a patched overlayer of thickness $t_{\mathrm{O(k)}}$ ($k = 2, 4$) and coverage $\Gamma$ ($0<\Gamma<1$), which indicates the fraction of surface covered by the overlayer. $R_2$ and $R_4$ data in Figure \ref{SM:fig:xps-O1s}d,e are fitted with the fitting parameter $\Gamma$, while $t_{\mathrm{O(k)}}$ is varied in \SI{1}{\angstrom} steps to obtain the best fit with $R^2 \approx 1$. On all samples, the adsorbates represented by component $\mathrm{O(4)}$ form a superficial layer of thickness $\approx$ \SI{4}{\angstrom}, corresponding to 1 monolayer \cite{domingo_water_2019}. In contrast, in the BTO/SRO/DSO sample, component $\mathrm{O(2)}$ is distributed below the BTO surface with thickness $\approx$ \SI{15}{\angstrom}, while in the BTO/SRO/GSO and BTO/SRO/SSO samples, component $\mathrm{O(2)}$ is concentrated near the surface with thickness $\approx$ \SI{6}{\angstrom}. The larger thickness $t_{\mathrm{O(2)}}$ in the BTO/SRO/DSO sample cannot be explained by a thicker overlayer of $\mathrm{O_LH^-}$ species above the BTO surface because the molecules in the overlayer would be in a different chemical environment and thus appear at a different binding energy. Instead, our experimental data suggest a scenario where $\mathrm{H^+}$ atoms are distributed below the BTO surface to form $\mathrm{O_LH^-}$ species (Figure \ref{fig:sketch}b), as already proposed in other works \cite{park_effect_2000, lee_imprint_2016}.

In summary, in the BTO/SRO/DSO sample, $\mathrm{O_LH^-}$ species have the largest distribution below the BTO surface and $\mathrm{H^+}$ atoms serve as charge compensation mechanism of the tail-to-tail polarization configuration revealed by XSW and PFM measurements. In the BTO/SRO/GSO sample, the comparable amount of $\mathrm{OH^-}$ and $\mathrm{O_LH^-}$ species near the surface favor a ferroelectric polarization with opposite orientation, resulting into a vanishing net polarization amplitude just below the BTO surface. Finally, in the BTO/SRO/SSO sample, the ferroelectric polarization is pointing upward in the unit cells near the surface with overall the largest amplitude among the samples under study. This is consistent with the largest amount of negatively charged hydroxyl or $\mathrm{O_2^-}$ molecules concentrated near the surface, as shown by component $\mathrm{O(4)}$ in Figure \ref{SM:fig:xps-O1s}c.

\subsection{XPS fit results}
\label{sec:xps-fit-results}

PE spectra were measured using a fixed mode of the electron analyzer with a pass energy of \SI{200}{\eV} [\SI{100}{\eV}] for Ba 4d, Ti 2p, C 1s [O 1s] core-level emission lines. In general, all PE spectra in this work were fitted using the software CasaXPS with Shirley background subtraction and a combination of Gaussian/Lorentzian functions with the best fit provided by ratio $70/30$ (for Ba 4d, O 1s, C 1s) and $40/60$ (for Ti 2p). A summary of the resulting BE shifts and FWHM of each component is reported in Table \ref{SM:tab:BE-shift_FWHM}.

\begin{table*}[h]
\caption{FWHM of each fit component and BE shift of components $\mathrm{P(m)}$ ($\mathrm{m=2, 3, 4, 5}$) from $\mathrm{P(1)}$ in Ba 4d, Ti 2p, O1s and C1s PE spectra.}

\begin{tabular}{p{1cm}p{1cm}p{1cm}p{1cm}p{1cm}p{1cm}p{1cm}p{1cm}p{1cm}}\hline
 & 
 \multicolumn{2}{c}{Ba 4d}  & \multicolumn{2}{c}{Ti 2p} & \multicolumn{2}{c}{O 1s}  & \multicolumn{2}{c}{C 1s}\\\hline
 \\[-1em]
  & BE shift (eV)& FWHM (eV)& BE shift (eV)& FWHM (eV)& BE shift (eV)& FWHM (eV)& BE shift (eV)& FWHM (eV)\\
 \hline
 $\mathrm{P(1)}$& \,\,\,\,-- &  0.83    & \,\,\,\,-- &   1.02   & \,\,\,\,-- &  1.14   & \,\,\,-- &  1.37  \\
 $\mathrm{P(2)}$& 2.58 &  0.83& 5.72 &  2.04&  1&    1.36& 1.4 & 1.75\\
 $\mathrm{P(3)}$& 1.21 &  1.65&      &      &  2&   1.32 &      &    \\
 $\mathrm{P(4)}$& 3.80 &  1.65&      &      &  2.7&  1.54&      &    \\
 $\mathrm{P(5)}$&      &      &      &      &  3.85&  1.45&      &    \\
 \hline
\end{tabular}
\label{SM:tab:BE-shift_FWHM}
\end{table*}

\subsection{\label{imfp}Inelastic mean free path}

The inelastic mean free path $\lambda_l(E_\nu)$ is defined, according to Ref. \cite{Shinotsuka18}, as: 
\begin{equation}
\lambda_l(E_\nu) = \frac{\mathcal{A}(E)(E_\nu- \mathrm{BE})}{E_p^2\bigl\{\mathcal{B}\left[ln(\mathcal{Y}\mathcal{A}(E_\nu)(E_\nu-\mathrm{BE})) \right]-\mathcal{C}/(E_\nu-\mathrm{BE})+\mathcal{D}/(E_\nu-\mathrm{BE})^2  \bigr\}}.
\label{SM:eq:lambda}
\end{equation}
In Equation \eqref{SM:eq:lambda}: $\mathcal{A}(E_\nu) = \left[1+(E_\nu-\mathrm{BE})/(2m_ec^2)\right]/\left[1+(E_\nu-\mathrm{BE})/(m_ec^2) \right]^2$, $\mathcal{B} = -1.0 + 9.44/(E_p^2+E_g^2)^{0.5}+0.69\rho^{0.1}$, $\mathcal{C} = 19.7-9.1\mathcal{U}$, $\mathcal{D} = 534-208\mathcal{U}$, $\mathcal{Y} = 0.191\rho^{-0.5}$, and $\mathcal{U} = N_v\rho/M$. In these equations: $\mathrm{BE}$ is the binding energy of photoelectrons from core level $l$, $m_e$ is the rest mass of electron, $c$ is the speed of light, $N_v$ is the total number of valence electrons per molecule, $\rho$ is bulk density, $M$ is the molecular weight, $E_p= 28.816( N_v.\rho/M )^{0.5}$ is the free-electron plasmon energy and $E_g$ is the band gap energy in eV. In the case of BTO: $N_v=24$, $M =$ \SI{233.19}{\g\per\mol} \cite{Patnaik02}, $\rho =$ \SI{6.02}{\g\per\cubic\cm} \cite{Patnaik02} and $E_g =$ \SI{3.76}{\eV} \cite{Kaveh12}.

\subsection{Photoelectron yield normalization}
\label{SM:sec:yield-normalization}

The PE yield undergoes two normalization steps: (i) by the incident X-ray intensity, and (ii) by the photoionization cross-section. First, the X-ray intensity $I_0$ is measured as the drain current from the last mirror before the sample. The incident X-ray intensity $I_0$ decreases by approximately $10\%$ in the energy range from $1400$ to \SI{1700}{\eV} (Figure \ref{SM:fig:normalization}b). This is due to the decreasing monochromator grating efficiency. Furthermore, the $I_0$ sawtooth profile in the region from $1480$ to \SI{1620}{\eV} results from the top-up electron injection at the Diamond Light Source. PE yield data normalized by $I_0$ are shown in Figure \ref{SM:fig:normalization}c. Here, the decrease in PE yield by a factor $\approx 1.8$ over the whole photon energy range follows from the varying photoionization cross-section \cite{Trzhaskovskaya01,Trzhaskovskaya02}. The second normalization step consists in dividing PE yield data by a second order polynomial resulting from the fit of $7$ points at each end of a yield curve in Figure \ref{SM:fig:normalization}c, where no XSW effect is observed. An example of fitting curve is shown in Figure \ref{SM:fig:normalization}c (blue solid line), while the normalized PE yield curves are reported in Figure \ref{SM:fig:normalization}d. Finally, at each photon energy $E_\nu$, average and standard deviation of PE yield data (measured under the same conditions) are calculated to determine the PE yield values and error bars reported in Figure \ref{fig:XSW-data}.

\begin{figure*}[t!]
\centering
\includegraphics[width=\textwidth]{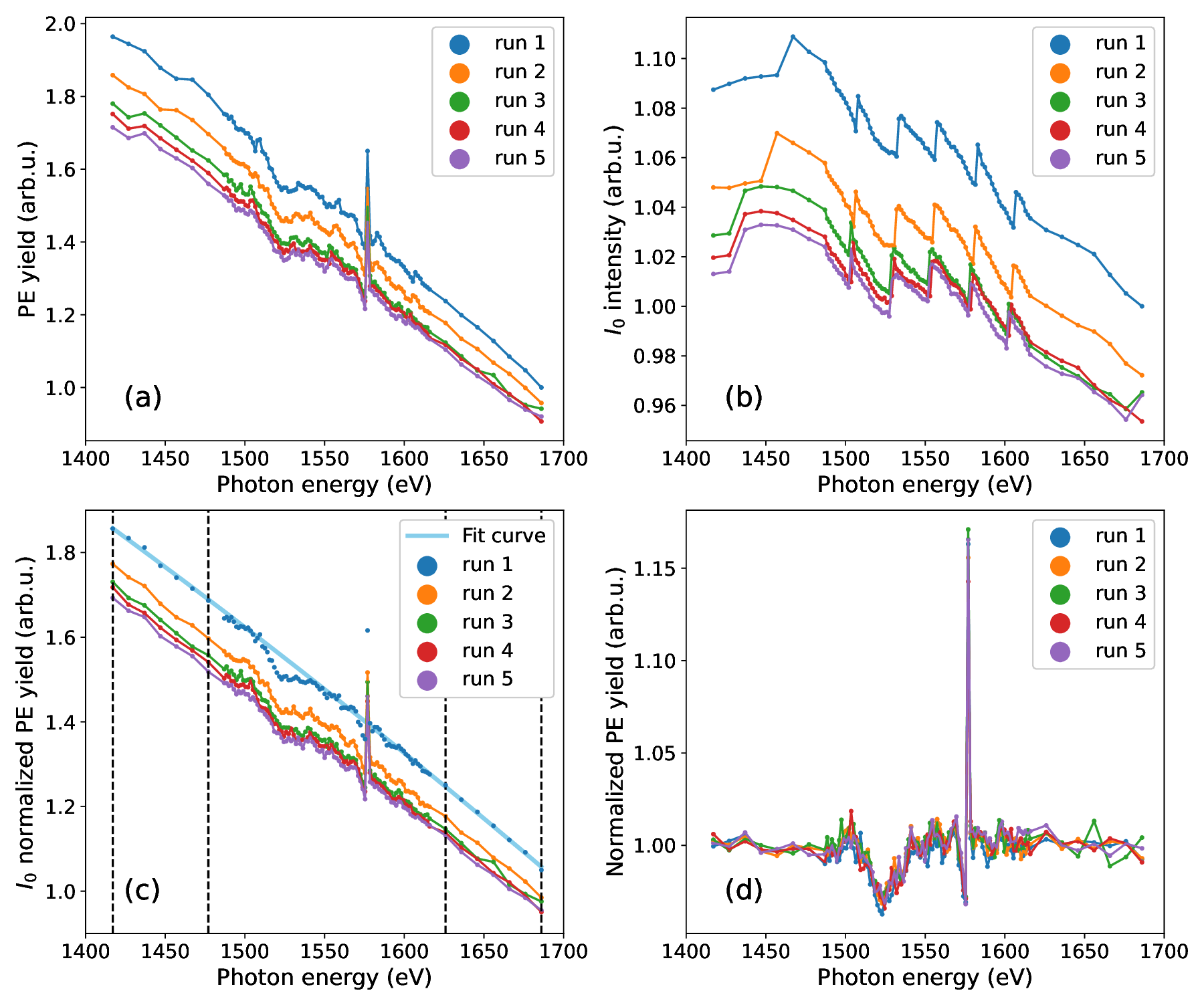}
\caption{(a) Ba 4d PE yield raw data of the BTO/SRO/DSO sample measured in $5$ consecutive photon energy scan (run). (b) Incident X-ray beam intensity $I_0$. (c) PE yield raw data normalized by $I_0$. (d) PE yield raw data normalized by $I_0$ and photoionization cross-section.}
\centering
\label{SM:fig:normalization}
\end{figure*}

\subsection{Reflection and transmission calculation}
\label{SM:sec:theory}

The X-ray diffracted intensity from the sublayer $L_i$ with thickness $t_i$ at photon energy $E_\nu$ and $z_i < z < z_i + t_i$ is calculated as \cite{Ivan01}:
\begin{equation}
  R(E_\nu, z) = \beta|Y|^2|r(E_\nu, z)|^2= \beta|Y|^2\Bigg|\frac{x_1-x_2 x_3 \exp(-\sigma \Delta z_i)}{1-x_3\exp(-\sigma \Delta z_i)}\Bigg|^2,
\label{SM:eq:R_z_en}
\end{equation}
where $x_1 = -\Big(b+\sqrt{b^2-C_1^2}\Big)/C_1$, $x_2 = -\Big(b-\sqrt{b^2-C_1^2}\Big)/C_1$, and $x_3 = (x_1-r_{t_i})\exp(\sigma t_i)/(x_2-r_{t_i})$, $\sigma = 2\mathrm{i}\sqrt{b^2-C_1^2}/L_{ex}$, and $\Delta z_i = z - z_i$. The reflection amplitude at the bottom [top] of $L_i$ is defined as $r_{t_i} = r(E_\nu, \Delta z_i = t_i)$ [$r(E_\nu, \Delta z_i = 0)$], and $r$ at the top of $L_i$ is then treated as the reflection at the bottom of layer $L_{i-1}$. Starting from the boundary condition of $r(E_\nu, z)=0$ at the bottom of the substrate, Equation \eqref{SM:eq:R_z_en} is employed recursively to calculate the diffracted intensity at the top of $L_0$, i.e. $R(E_\nu, 0)$ (Section \ref{sec:theory}).  

Parameters appearing in Equation \eqref{SM:eq:R_z_en} are summarized in the following. In particular, $b$ is defined as $ b = -y(E_\nu) - \mathrm{i}y_0 + y_\varphi(z)$ where $y(E_\nu) = 2\sqrt{\beta}(\sin^2{\theta_B})(E_\nu - E_B)/(E_B X_\mathrm{r}) + \chi_{0\mathrm{r}}(1+\beta)/(2\sqrt{\beta}X_\mathrm{r})$, $y_0 = (\chi_{0\mathrm{i}}(1+\beta))/(2\sqrt{\beta}X_\mathrm{r})$,
$y_\varphi(z) = (L_{ex}/2) d\varphi(z)/dz$. Here, $y(E_\nu)$ is a dimensionless parameter that indicates the energy deviation from the exact Bragg energy $E_B$ during incident photon energy scan, $y_0$ represents the attenuation of X-ray intensity due to photoelectric absorption and $y_\varphi(z)$ indicates the shift of diffraction planes due to lattice deformation. In the latter equations, the geometry factor $\beta$ is defined as $\beta = \Gamma_{\boldsymbol{0}} / |\Gamma_{\boldsymbol{h}}|$, where $\Gamma_{\boldsymbol{0}} = k_{\boldsymbol{0}z}/K$ and $\Gamma_{\boldsymbol{h}}=k_{\boldsymbol{h}z}/K$ are the direction parameters with $K=2\pi/\lambda_B$ and $\lambda_B$ is the Bragg wavelength. The extinction length represents the penetration depth of the XSW field and is defined as $L_{ex} = (\lambda_B \Gamma_{\boldsymbol{0}})/(\pi \sqrt{\beta}X_\mathrm{r})$.

The parameter $C_1$ has the form $C_1=C(1-\mathrm{i}p)\exp(-W(z))$, where $p=-X_\mathrm{i}/X_\mathrm{r}$ and $C$ is the polarization factor which is equal to 1 for $\sigma$ polarization and $\cos{2\theta_B}$ for $\pi$ polarization, with $\theta_B$ being the Bragg angle. The parameters $Y = \sqrt{\chi_{\boldsymbol{h}}/\chi_{\Bar{\boldsymbol{h}}}} = |Y|\exp(\mathrm{i}\Phi_Y)$, $X_{\mathrm{r}}=Re\Big[\sqrt{\chi_{\boldsymbol{h}} \chi_{\Bar{\boldsymbol{h}}}}\Big]$ and $X_\mathrm{i}=Im\Big[\sqrt{\chi_{\boldsymbol{h}} \chi_{\Bar{\boldsymbol{h}}}}\Big]$ are derived from crystal susceptibilities $\chi_{0}$, $\chi_{\boldsymbol{h}}$, $\chi_{\Bar{\boldsymbol{h}}}$ corresponding to vectors $\boldsymbol{0}*\boldsymbol{h}$, $\boldsymbol{h}$, and $-\boldsymbol{h}$ with $\boldsymbol{h} = 2\pi\boldsymbol{H}$. The crystal susceptibility $\chi_{\boldsymbol{h}}$ is generally a complex number $\chi_{\boldsymbol{h}} = \chi_{\boldsymbol{h}\mathrm{r}} + \mathrm{i}\chi_{\boldsymbol{h}\mathrm{i}}$, where $\chi_{\boldsymbol{h}\mathrm{r}} = -\Big(\mathrm{e}^2\lambda_B^2/mc^2\pi \Omega\Big)F_{\boldsymbol{h}\mathrm{r}}$ represents X-ray elastic scattering, while $\chi_{\boldsymbol{h}\mathrm{i}} = \Big(\mathrm{e}^2\lambda_B^2/mc^2\pi \Omega\Big)F_{\boldsymbol{h}\mathrm{i}}$ stands for X-ray absorption. The structure factor $F_\mathrm{h} = \sum_j f_j \exp(-W_j) \exp(-\mathrm{i}\boldsymbol{h} \boldsymbol{\rho}_j)$ is computed from the atomic scattering factor $f_j$ of $j$th atom at position vector $\boldsymbol{\rho}_j$ in the unit cell. The atomic scattering factor $f_j = f_0(\theta_B, \lambda_B, Z) - Z + f_1(0, \lambda_B, Z) + \mathrm{i}f_2(0, \lambda_B, Z)$ describes the interaction between X-rays and atoms, where $Z$ is the atomic number, $f_0$, $f_1$, and $f_2$ are tabulated in Ref.~\cite{Maslen06, Henke93}.

Finally, besides the reflection amplitude, solving the Takagi-Taupin equations also provides the transmission amplitude through the sublayer $L_i$ \cite{Ivan01}:
\begin{equation}
T(E_\nu, z) = \exp[\mathrm{i}\Phi(E_\nu)\Delta z_i/2]\Bigg(\frac{1-x_3\exp(-\sigma \Delta z_i)}{1-x_3} \Bigg), 
\label{SM:eq:T}
\end{equation}
where $\Phi(E_\nu)=(2\pi\chi_{\boldsymbol{0}})/(\lambda_B \Gamma_{\boldsymbol{0}})- 2C_1 x_1/L_{ex}$.

\subsection{Deformation phase calculation}
\label{SM:sec:deformation_phase}

The deformation phase $\varphi_0$ in Equation 3 in the article, is derived from the bicrystal model \cite{Ivan01} and is defined as $\varphi_0 = 2 \pi ( c_0 - \overline{c})t_0/\overline{c}^2$. The bicrystal model assumes: a crystal, in which the XSW is formed, and a deformed overlayer of thickness $t_l$ that induces a shift of the diffraction planes $\varphi_0 = 2 \pi ( c_l - c_c)t_l/c_c^2$, where $c_l$ and $c_c$ are the out-of-plane lattice parameters of the deformed layer and the crystal, respectively. In our samples, the XSW forms in the BTO film itself with a deformation given by an inhomogeneous strain. Therefore the deformation phase cannot refer to the $c$ parameter of a single crystal underneath. Instead, $\varphi_0$ refers to the average out-of-plane lattice parameter $\overline{c}$, calculated from the experimental diffraction curves (Section \ref{SM:sec:c_average}). This is equivalent to modelling the thin film as a crystal with an average out-of-plane parameter $\overline{c}$ and an increasing [decreasing] deformation phase towards the interface with SRO [towards the surface]. The validity of this choice is confirmed by the coherent position of the Ba atoms (Table II) being close to $1$, i.e., near the diffraction planes. Conversely, referring, e.g., the deformation phase to the bottom sublayer $L_{n-1}$ would lead at the top unit cells to unphysical positions of the Ba atoms \SI{1}{\angstrom} away from the diffraction planes.

\subsection{The uncertainty of Ba and Ti atomic positions}
\label{SM:sec:uncertainty}

The attainable structural accuracy in the determination of the average atomic positions depends ultimately on the error bar of the PE yield profiles according to Poisson statistics. We observe that the error bars of data of the BTO/SRO/SSO sample ($1\% < \sigma_\kappa < 7\%$) are generally larger than those of the other two samples ($\sigma_\kappa < 2\%$). This is due to the greater amount of adsorbates on the BTO/SRO/SSO surface, which attenuated the measured PE intensity (Section \ref{SM:sec:Ti-Ba-C-XPS}). Besides, lower PE intensity is also expected with a decreasing $\gamma$ or smaller $\lambda_{l,\gamma}$. This explains the trend of increasing $\sigma_\kappa$ with decreasing $\gamma$ and the larger uncertainty in Ti positions compared to Ba (Table \ref{tab:Pc-Fc}). The latter observation is also due to the smaller photoionization cross-section of Ti 2p as compared to Ba 4d core levels \cite{Trzhaskovskaya01,Trzhaskovskaya02}.

\end{justify}

\end{document}